\documentclass[preprint,preprintnumbers,amsmath,amssymb]{revtex4}
\pdfoutput=1
\usepackage{graphicx}

\begin{document}

\title{Enhanced imaging of microcalcifications in digital breast
tomosynthesis through improved image-reconstruction algorithms
}

\author{Emil Y. Sidky}
\email{sidky@uchicago.edu}

\author{Xiaochuan Pan}
\email{xpan@uchicago.edu}

\author{Ingrid S. Reiser}
%\email{ireiser@uchicago.edu}

\author{Robert M. Nishikawa}
%\email{r-nishikawa@uchicago.edu}
\affiliation{%
University of Chicago \\
Department of Radiology \\
5841 S. Maryland Ave., Chicago IL, 60637
}%

\author{Richard H. Moore}

\author{Daniel B. Kopans}
\affiliation{%
Massachusetts General Hospital \\
Boston, Massachusetts 02114
}%

\date{\today}% It is always \today, today,

\begin{abstract}
PURPOSE:\\
\noindent
We develop a practical, iterative algorithm for
image-reconstruction in under-sampled tomographic
systems, such as digital breast tomosynthesis (DBT).

METHOD:\\
\noindent
The algorithm controls image regularity by minimizing
the image total $p$-variation (TpV), a function that reduces to
the total variation when $p=1.0$ or the image roughness
when $p=2.0$.  Constraints on the image, such as image positivity
and estimated projection-data tolerance, are enforced by
projection onto convex sets (POCS). The fact that the tomographic
system is under-sampled translates to the mathematical property
that many widely varied resultant volumes may correspond to a given data
tolerance. Thus the application of image regularity serves
two purposes: (1) reduction of the number of resultant volumes
out of those allowed
by fixing the data tolerance, finding the minimum image TpV for
fixed data tolerance, and (2) traditional regularization, sacrificing
data fidelity for higher image regularity. The present algorithm
allows for this dual role of image regularity in under-sampled tomography.

RESULTS:\\
\noindent
The proposed image-reconstruction algorithm is applied to three clinical DBT data sets.
The DBT cases include one with
microcalcifications and two with masses.

CONCLUSION:\\
\noindent
Results indicate that there
may be a substantial advantage in using the present image-reconstruction
algorithm for microcalcification imaging.
\end{abstract}

%\pacs{Don't forget me}

\maketitle

\section{Introduction}
\label{sec:intro}

Digital breast tomosynthesis (DBT) is an emerging X-ray imaging modality that
aims at improving the effectiveness of mammographic screening without
an increase in radiation dose.  DBT provides partial tomographic
information that aids in reducing the impact of overlapping tissue structures
on tumor detection \cite{Niklason:1997,Dobbins:2003}. 
A key component of the system is the image-reconstruction (or synthesis) algorithm.
Data acquired in DBT are far from sufficient for ``exact'' tomographic
image-reconstruction, which limit the effectiveness of single-pass algorithms.
Such algorithms are generally derived from algorithms that assume complete
tomographic data, and they generally introduce artifacts in the DBT images.
Nonetheless, one-pass algorithms such as
filtered back-projection (FBP), modified FBP and matrix-inversion methods
are employed to produce images.
A thorough investigation on DBT image
reconstruction algorithms \cite{Wu:2004,Wu:2004RSNA,Zhang:2006}, 
showed that iterative algorithms present many advantages
over one-pass algorithms.  Reasons for this include (1)
iterative algorithms
generally put milder assumptions on the ``missing'' data; most FBP algorithms
set missing views to zero -- which is an impossibility for projection imaging,
and (2) iterative algorithms allow for physical constraints to be easily incorporated
such as physical borders of the object, and valid range for X-ray attenuation values.
Here, we investigate
iterative image-reconstruction in DBT based on image
total p-variation (TpV) minimization \cite{chartrand-2007-exact,Sidky-MIC-tpv:07}.

Investigation of existing iterative algorithms applied to DBT has been 
performed in Refs. \cite{Wu:2004,Wu:2004RSNA,Zhang:2006}. These references cover the principal
iterative algorithms used in tomographic image-reconstruction, demonstrating
their performance on various imaging features pertaining to DBT.  Maximum likelihood (ML)
methods and variations on the algebraic reconstruction technique (ART) are studied.
These iterative algorithms, however, may not be ideally suited to image-reconstruction
in DBT.  Generally speaking, iterative algorithms have been designed to
work efficiently for
scanning systems where the projection data are complete, or nearly
complete, but of low quality.
For example, in most nuclear medicine imaging systems, % \cite{??},
the collected projection data is usually
fully sampled allowing for ``exact'' inversion, at least theoretically, but the
data are often corrupted by high levels of noise. As a result, an iterative
algorithm is often employed. 
DBT scanning is challenging for image-reconstruction algorithms in a different way.
The data are of high quality (low noise), but they are radically incomplete.
This incompleteness means that there may be many, very different, candidate
attenuation distributions that
agree with the available data. In fact, the recent interest in compressive
sensing \cite{candes-2006-robust,candes-2006-stable},
poses the extreme limit of the latter
situation: namely, can one obtain exact image-reconstruction from ``perfect''
quality data that is under-sampled.  In this article, we adapt an algorithm \cite{Sidky-CCB:08},
which we have developed for investigating compressive sensing in tomographic
image-reconstruction, to the DBT scanning system.

Iterative image-reconstruction algorithms aim to minimize an objective
function that combines a data fidelity term and a regularization term.
The
overall picture is that there is a trade-off between the two terms.
When the weight on the regularization term is small the resulting image
yields data that is ``close'' to the available data, but it may
contain conspicuous artifacts due to noise or other inconsistencies in the
data. When the weight on the regularization term is large, the resulting
image will be regularized at the expense of faithfulness to the data.
This picture applies to the scanning situation where the data are complete,
but of low quality.
For incomplete data scans, however, this trade-off picture is too simple.
One of the basic properties of a tomographic system that collects incomplete
projection data is that there is not a unique image that corresponds to the
available projection data. As a result,
regularization of the image takes on two roles:
(1) selection of a unique image among those that agree with the projection
data, and (2) the traditional role where the image is regularized while relaxing
consistency with the available data.  In the first role, the image regularization
is lowered while the image is constrained to a given data agreement. In the
second role the data constraint on the image is relaxed allowing for further
minimization of the image regularization.

In our previous work, the image
reconstruction algorithm employed projection onto convex sets (POCS) to enforce
a data consistency constraint as well as other physical constraints such
as positivity, and steepest descent was used to minimize the regularization term.
There was an adaptive element introduced to control the relative step-sizes
of the POCS and steepest descent components of the algorithm, hence the algorithm
is called adaptive steepest descent - POCS (ASD-POCS) \cite{Sidky-CCB:08}. The ASD-POCS
algorithm allows for the separation of the two roles for the regularizer in tomographic
image-reconstruction from incomplete projection-data.  Our previous
work was focused on compressive sensing in tomography and was restricted to
$\ell_1$-based regularizers, and algorithm efficiency
was a secondary concern.
% Here, we aim at practical image-reconstruction for DBT,
%and some modifications are made to the original ASD-POCS algorithm to allow for
%useful images at low iteration numbers. 

In this article, we break-up the pieces of the ASD-POCS algorithm, and reassemble
them into a simplified, practical image-reconstruction algorithm that we
apply to DBT.  The practical aspect refers to the fact that we aim to
obtain useful images within 10-20 iterations, and the simplification of
the algorithm refers to a reduction in the number of algorithm parameters
to only those that have a significant impact on the image within the first
few iteration steps. Although we provide a specific algorithm here, we do
not claim that it is optimal; there are likely many ways to reassemble
the ASD-POCS algorithm pieces that yield useful tomographic images. As a result,
we refer to ASD-POCS as a framework instead of a single algorithm.
Few quantitative comparisons are made
as such detailed comparisons make sense only when a particular scan geometry,
set of reconstruction parameters, and image regularizer is selected.
The various images are shown to reveal the effect of various algorithm
parameters on the reconstructed images.

The remainder of the
paper is organized as follows: Sec. \ref{sec:DBTsys} describes the
general data model for iterative image-reconstruction in X-ray based
tomography,
Sec. \ref{sec:current} motivates the need for a new type of iterative algorithm
for incomplete scanning configurations such as DBT,
Sec. \ref{sec:ASD-POCS} presents an image-reconstruction algorithm for DBT
derived within the ASD-POCS framework, and
Sec. \ref{sec:results} demonstrates the image-reconstruction algorithm with
actual DBT case data that contains both microcalcifications and masses.

\section{System model and image-reconstruction}
\label{sec:DBTsys}

We describe the system model for X-ray tomography for which we develop
the image-reconstruction algorithm from the ASD-POCS framework.
On the one hand, the presentation is quite general in that the image-reconstruction
algorithm can be applied to a wide class of linear system models. On the
other hand, many aspects of the algorithm implementation are quite specific.
For example, the representation of the imaging volume, i.e. voxel shape, is
designed with the DBT scan in mind.  In this introductory section, we
aim the discussion toward general X-ray tomography, but we specify the particular
geometry and implementations used here to obtain the DBT results.

%In past applications of iterative image reconstruction, the emphasis has been primarily
%on systems that provide theoretically (nearly) complete projection data that contain a
%high level of noise. For example, iterative image reconstruction has found application
%in nuclear medicine imaging \cite{NMI}, and potential application in low-dose CT \cite{CT}.
%In such cases the algorithm's performance does not in general depend strongly on how the
%data model is computed. For example, the particular algorithm used for computing
%line integrals through the imaging volume is not as important as, say, the form
%of the penalty function or what physical factors are encorporated into the model.
%For the DBT scanning configuration, the data sampling is far from complete, and as
%a result, the reconstructed image may depend strongly on both what is included in the
%data model and how the model is computed.  The impact of this strong dependence is that
%it is crucial to specify every aspect of the data model which the algorithm seeks to
%invert, and that it is in general not possible to equate performances of seemingly
%similar algorithms.  The result of this fact is that we take great care to specify
%all conditions that yield the images of Sec. \ref{sec:results}, including the system
%configuration and model formulation in this section,
%and any comparisons made are of a qualitative nature.

\begin{figure}[ht]

\begin{minipage}[b]{0.8\linewidth}
\centering
\centerline{\includegraphics[width=10cm,clip=TRUE]{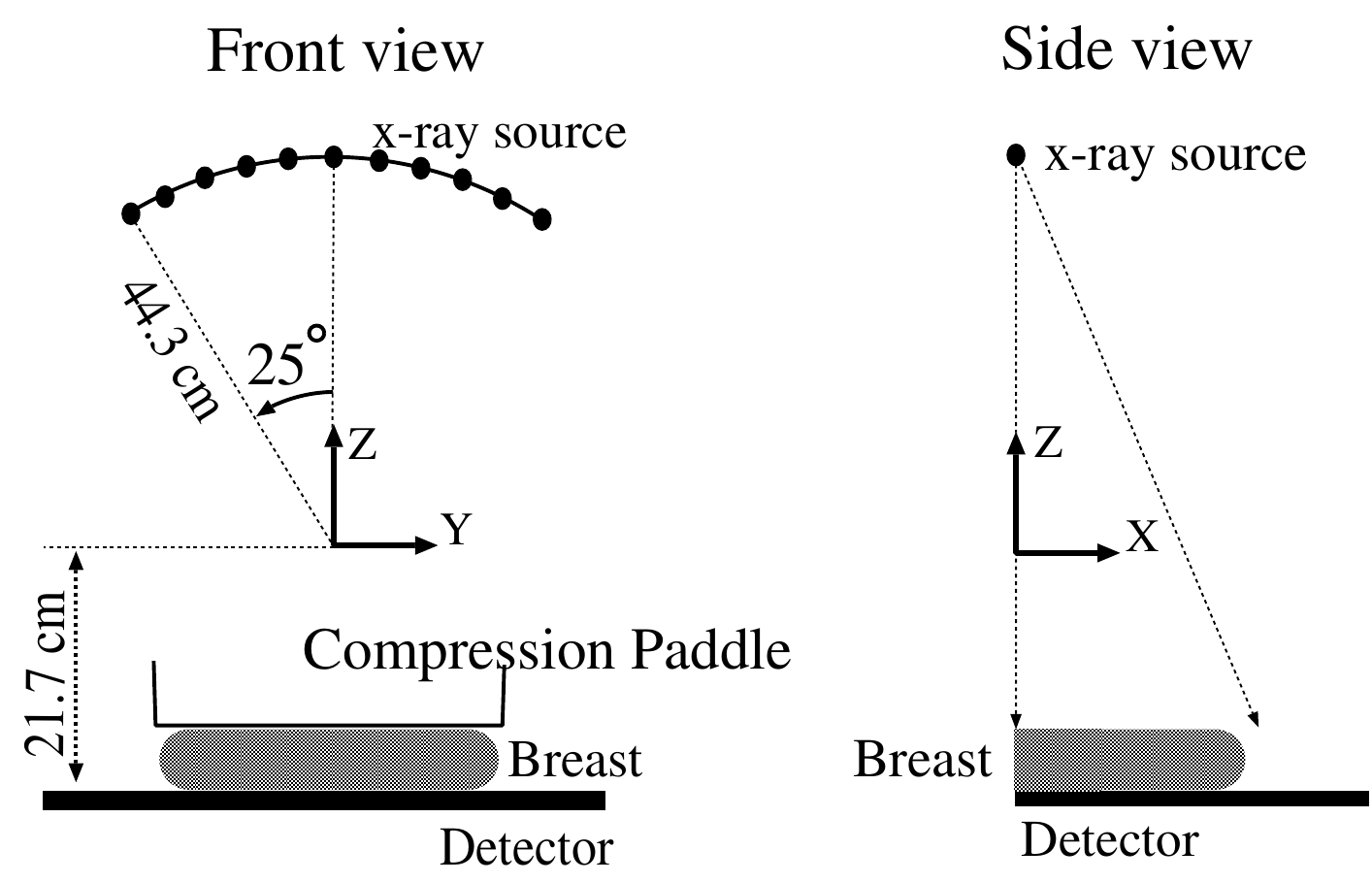}}
\end{minipage}
\caption{
Configuration of the digital breast tomosynthesis system. The coordinate
system, whose origin lies on the center of rotation for the X-ray source,
is also indicated. The front view shows a schematic including the compression paddle.
The walls of this paddle are visible in many of the projections.
\label{fig:DBTscanner}}
\end{figure}

DBT has undergone much development recently, and there are
two main configurations being pursued.
% General Electric, Siemens, Phillips, Sectra, Planmed, and Hologic 
Most companies working on DBT are developing variations of a swinging X-ray
source, while XCounter is proposing a linear X-ray movement system.
The common denominator for DBT systems
is that projection data is acquired over a limited number of angles 
with respect to a full, circular tomographic scan as acquired in CT.  For the present study
we perform volume reconstruction from data acquired by a DBT prototype
developed at Massachusetts General Hospital in collaboration with General Electric
Healthcare.  The scanner configuration and properties are specified
in Ref. \cite{Wu:2004}, but we re-iterate the geometric configuration here.
As shown in Fig. \ref{fig:DBTscanner}, the breast is compressed to a thickness of
3-8 cm on a carbon-fiber tray protecting the
fixed, flat-panel detector. The X-ray source is moved on an arc, centered
on point $h=21.7$ cm above the detector, and with radius $R=44.3$ cm. The detector
is composed of an array of 1800x2304 detector bins with width 100 microns, and
is physical dimensions are $W=180.0$ mm  $\times$  $L=230.4$ mm. The number of projections
is 11, and they are approximately
equally spaced along the 50$^\circ$ arc. In the article we
use the term
"in-plane" to refer to $xy$-planes, parallel
to the detector, and the term "depth" to refer to the $z$-direction, perpendicular to the
detector.

The data at each detector bin can be approximately related to the line integral
of the breast X-ray attenuation-map:
\begin{equation}
\label{dataModel}
g(s,u,v) = \int d \ell f(\vec{r}_0(s) + \ell \hat{\theta}(s,u,v)),
\end{equation}
where the source position follows
\begin{equation}
\vec{r}_0(s) = (0, R \sin{s}, R \cos{s}),
\end{equation}
and the detector bin locations are described by
\begin{equation}
\vec{d}(u,v) = (u , v - L/2 , -h).
\end{equation}
The unit vector $\hat{\theta}(s,u,v)$ points from X-ray source to
detector bin:
\begin{equation}
\hat{\theta}(s,u,v) = \frac{\vec{d}(u,v)-\vec{r}_0(s)}{|\vec{d}(u,v)-\vec{r}_0(s)|}.
\end{equation}
The data model in Eq. (\ref{dataModel}) involves integration of the continuous
object. But for the majority of iterative image-reconstruction algorithms further
approximation is necessary, because these algorithms generally apply to only
finite linear systems and as a result the imaging volume must have a finite
representation.

%In deriving analytic inversion formulas for x-ray tomography, the data model
%is generally taken to be line-integrals of the x-ray attenuation map:
%\begin{equation}
%\label{dataModel}
%g(\vec{r}_0,\hat{\theta}) = \int d \ell f(\vec{r}_0 + \ell \hat{theta}),
%\end{equation}
%where $\vec{r}_0$ is the x-ray source location, and $\hat{theta}$ indicates
%the direction from the x-ray source to a particular bin on the detector.
For the discussion below this imaging equation is converted to a discrete,
linear system:
\begin{equation}
\label{linsys}
M \vec{f} = \tilde{g}.
\end{equation}
The image vector, $\vec{f}$, is a finite set of coefficients specifying the
particular combination of basis elements, which in this case are voxels.
The available set of projection data, $\tilde{g}$, will in general have
a different size than the set of image basis elements.  The system
matrix $M$ approximates the continuous line integration of Eq. (\ref{dataModel}).
The particular form of $M$ depends on how the integration approximation
is formulated and on the choice of image basis functions.
For the current work, we employ the standard voxel representation of the imaging
volume.
% We point out, however, that alternative representations have been
%explored in the literature \cite{Yagel,Muller,Ziegler}.
The choice of voxel dimensions typical
in DBT are asymmetric.
For specifying the voxel size, the in-plane 
resolution is taken to be the detector resolution -- in this case 100 microns. The
depth resolution, however, is about 10-fold lower.  In previous work, the voxel size has been
taken as 0.1x0.1x1.0 mm$^3$ \cite{Wu:2003,Wu:2004}, and we do the same.
% As will
%be apparent in the subsequent sections, this assymetric choice of voxel size has
%quite some impact on image reconstruction.
With this choice of voxel dimension, the
imaging volume is composed of 30 to 80 slices arranged parallel to the detector
and within each slice there are the same number of voxels as detector bins.
For the reconstructions presented in the results, the slice number is fixed
at 60.

Before going on to specify the exact form of $M$, we take an aside here to
discuss projection data incompleteness. The important point about incomplete
scanning data, is that there may be many attenuation distributions
that agree with the available
projection data, or equivalently, that solve Eq. (\ref{linsys}).  There are two
aspects to the data incompleteness: the number of measurements may be less
than the number of unknowns, and the system matrix $M$ may be ill-conditioned.
DBT suffers from both types of incompleteness.  For the present imaging
volume the number of unknown voxel values is 110,880,000, while the number of measured
rays for the 11-projection DBT data set is 22,351,560. Thus, based on vector
dimensions alone, the DBT system is under-sampled by a factor of 5.
A way to think about the stability
issue is that there may be many attenuation distributions
that approximately solve Eq. (\ref{linsys}),
or more precisely, given a ``small'', positive number $\delta$ many images
may satisfy the following inequality:
\begin{equation}
\label{linsysStability}
\| M \vec{f}-\tilde{g} \| \le \delta.
\end{equation}
For example, if the number of views is increased by a factor of 10
the DBT system may still suffer from the second kind of data
incompleteness because the geometrical arrangement of the measured rays
may not be optimal for tomographic image-reconstruction.
The incompleteness in the DBT
scan means small changes in the reconstruction algorithm may have a large effect
on the reconstructed images, and the data incompleteness plays an integral role
in the algorithm design of Sec. \ref{sec:ASD-POCS}.

%(?? Should we put the exact equations for M ??)
The projection matrix $M$ employed here is ray-driven, meaning that the individual
rays of the projection are first identified and the contribution of image
voxels to the individual rays is computed. For each ray in the projection data set,
the intersection of that ray with the mid-plane of each slice is computed.
%FORMULAS.
The contribution of the ray-integral for a particular slice is obtained
by linearly interpolating
the neighboring four voxel values within the slice and multiplying the result
by the ray path-length through the slice.
%FORMULAS.
Each of the slice contributions are subsequently summed to yield the ray integral.
%SUM.
In practice, the size of $M$ is enormous. For the present set-up using 60 slices,
$M$ has on the order of $10^{15}$ elements.  Typically, $M$ is computed on-the-fly which is
quite efficient for projection,  because at most 240 voxels contribute to each
ray integration.

The above discussion specifies the form of the linear system that we seek to solve.
In the next section, the need for a new algorithm is motivated.
% by displaying
%results from basic forms of existing iterative algorithms such
%projection onto convex sets (POCS) and expectation-maximization
%(EM).

\section{Iterative algorithms and DBT image-reconstruction}
\label{sec:current}

As we have discussed above, the DBT scanning system yields
incomplete data for tomographic image-reconstruction.
% For this type
%of system, where the projection data are of high quality
%yet under-sampled, we motivate the need for the ASD-POCS framework.
Most of the commonly used iterative algorithms are based on an
optimization problem containing two terms: (1) data error $\delta$, the difference
between the available data and the estimated projection data based
on the current image estimate, and (2) an image regularity penalty,
some function, $R(\cdot)$,
of the image that increases with ``roughness'' or
some other undesirable property of the image.  The function $R(\cdot)$
can take many forms, such as image total variation or squared
voxel differences, the roughness. The data-error, can also take
different functional forms. The usual optimization problem
minimizes an objective function that is the sum of these
two terms combined with a parameter to control the strength
of the regularization. 

\begin{figure}[ht]

\begin{minipage}[b]{0.8\linewidth}
\centering
\centerline{\includegraphics[width=6cm,clip=TRUE]{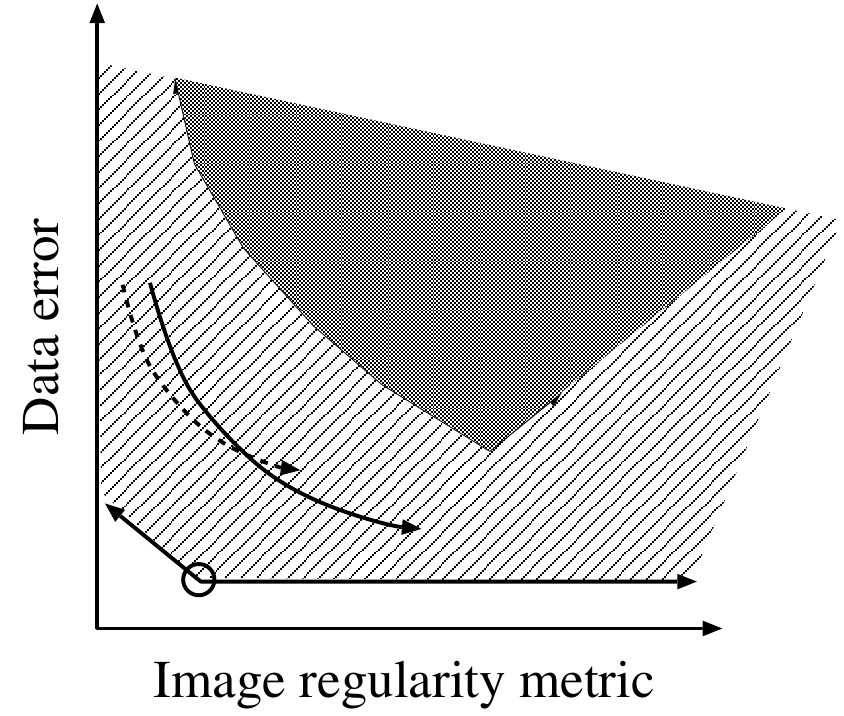}}
\end{minipage}
\caption{Diagram of in the $R,\delta$-plane comparing possible
images for an under-sampled versus a completely-sampled tomographic system.
The dark region represents images of the latter case.  For completely-sampled
systems, a unique image minimizes the data error, $\delta$, hence only
one value of $R$ is possible. For under-sampled systems, the lightly shaded
region, many possible candidate volumes
correspond to the situation of minimum $\delta$.  The circled point has
significance for compressive sensing, if $R$ is the $\ell_1$-norm 
or image total-variation.  The two curves represent generic
behavior of standard iterative algorithms for the case of no regularization
(solid curve), and with regularization (dashed curve).
\label{fig:trajSketch}}
\end{figure}

The sketch in Fig. \ref{fig:trajSketch} illustrates the difference
between the present
DBT scanning system and tomographic systems with complete but
low-quality data. Each point on the 
$R,\delta$-plane represents an image estimate, or possibly multiple image
estimates corresponding to the same data-error and image-regularity measure.
The dark-shaded region is indicative of completely-sampled,
high-noise system.  Lower values of the data-error generally
leads to worse image-regularity. Minimizing the data-error leads
to a very small set (possibly only one) of image estimates that
is generally very noisy. Hence, rarely is the image at the bottom
of the dark region, minimum $\delta$ ever sought. Instead, a regularity penalty
is introduced in to the objective function,
and an image along the left-edge of the dark
region is obtained, yielding a smoother image with greater data-error.
The light-shaded region represents possible image estimates
for an under-sampled, low-noise scanning system, such as DBT.
The achievable data-errors are much lower because the data are
of higher quality and there is generally less inconsistency
when the projection data are under-sampled. As the system is
under-sampled, there is not a unique image that minimizes
the data error. In the schematic, there may be many images
with different values of the regularity measure that have the
minimum data error. As a result, for an effective image
reconstruction algorithm for undersampled tomographic
systems, it is desirable to be able to independently
control the data error and regularity of the image estimates.

The curves shown in Fig. \ref{fig:trajSketch} sketch possible
trajectories of standard iterative methods applied to the
under-sampled system.  The solid curve represents iterations
from a generic algorithm that minimizes data error. If the
algorithm is initialized with a uniform image, as is often done,
the image regularity measure starts at low values and the data
error is high. As the iterations progress, the image estimate
migrates down and to the right. Reduced data error is obtained,
generally, at the expense of worse image regularity. If a
penalty term is introduced, one might obtain the dashed curve.
The image estimates will have lower values of $R(\cdot)$,
but the data-error will decrease more slowly. As a
result, iterative algorithms that include a penalty term of
fixed strength may not be the most efficient for under-sampled
tomographic image-reconstruction.

The ASD-POCS algorithm, we developed in Ref. \cite{Sidky-CCB:08},
was designed for compressive-sensing tomographic image-reconstruction.
Specifically, it was designed to solve the following
constrained minimization
\begin{equation}
\label{rmin}
\vec{f}^* = \text{argmin} \; R(\vec{f}),
\end{equation}
subject to the constraints
\begin{align}
\label{constraints}
& \left| M \vec{f} - \tilde{g} \right| ^2   \le \epsilon^2, \\
& \vec{f}  \ge 0. \notag
\end{align}
For the compressive sensing application,
the ASD-POCS algorithm uses the image total
variation (TV) as the regularity measure $R(\cdot)$. The minimum
TV image is sought for a fixed data error $\epsilon$ ($\delta \le \epsilon$).
Minimum TV images have the
sparsest gradient magnitude images, which is an assumption that
applies well to underlying images that are piecewise constant.
In particular, one of the goals of ASD-POCS is to closely approximate
the image with minimum data error and minimum TV, indicated by
the circle in Fig. \ref{fig:trajSketch}.  More generally, the
ASD-POCS algorithm can be used to search the lightly-shaded
region of the figure, and the function $R(\cdot)$ may take other
forms.

\section{a practical image-reconstruction algorithm using the ASD-POCS framework}
\label{sec:ASD-POCS}

Although the ASD-POCS algorithm is effective at finding a close
approximation to the solution of the constrained minimization Eqs. (\ref{rmin})
and (\ref{constraints}), it may take hundreds to thousands of iterations
to obtain a satisfactory solution.  Keeping practicality in mind, we
assemble an algorithm within the
ASD-POCS framework that is  more efficient and employs
fewer algorithm parameters.

The ASD-POCS algorithm solves the constrained minimization problem by employing
POCS to enforce the convex constraints on the image combined with steepest descent
to reduce the $R(\cdot)$ objective function. 
One modification
is that we include a line
search on the steepest descent portion of the algorithm. The line search ensures
that the steepest-descent steps actually reduce the objective $R(\cdot)$ from the first
iteration on. This change reduces artifacts in the early iterations (this is not done
in the original ASD-POCS algorithm, because it may sacrifice
the ability of the algorithm to yield a good approximation to the constrained-minimization
problem). Another important modification
is reducing the number of control parameters for the adaptation
of the step-sizes.  The previous version of ASD-POCS had 6 control parameters,
which served its purpose of obtaining a good approximate solution to the constrained
minimization problem. Because the optimization problem, Eqs. (\ref{rmin}) and (\ref{constraints}),
was being solved, the 6 control parameters affect only the ``path'' of the image estimate
but the final image could be regarded as depending only on the single
parameter $\epsilon$ in the constraint.
For the present case, where we intend to truncate the iteration well short of convergence,
the reconstructed image has to be viewed as a function of the algorithm parameters and $\epsilon$.
Having to explore the impact of seven parameters negates the advantage of truncating
the iteration early.

We present the new version of the ASD-POCS algorithm in the
form of a pseudo-code and abbreviate the notation where possible.
The symbol $:=$ means assignment, meaning that the result on the right-hand side gets assigned to
the variable on the left-hand side; image-space variables have a vector sign, e.g.
$\vec{f}$, and a hat is used if the vector has unit length;
data-space variables are denoted by a tilde, e.g. $\tilde{g}$.
The number of measured rays, length of $\tilde{g}$, is $N_d$.
The vector $\vec{M}_i$ is the row of the system matrix that yields the
$i$th data element. The function $P$ enforces lower and upper bounds on
an image estimate: $P(\vec{f},a,b)$  yields the image $\vec{f}^\prime$
with components
\begin{equation}
f^\prime_i =
\begin{cases}
a & f_i< a \\
f_i & a \le f_i \le b \\
b & f_i>b
\end{cases}
.
\notag
\end{equation}
The function $R(\cdot)$ is the image regularity measure.

\noindent
The pseudo-code is:

\begin{tt}
\begin{tabbing}
~~~~~~~\=~~~\=~~~\=~~~~~\=~~~~~~~~~~~~~~~~~~~~~~~~~~~~~\=~~~~~\=\\
\noindent
1:  \> $\beta := 1.0; \; N_\text{iter}=10$ \\
2:  \> $ng := 5$\\
3:  \> $r_\text{max} := 1.0$ \\
4:  \> $\gamma_\text{red}:=0.8$ \\
5:  \> $\vec{f} := 0$ \\
6:  \> for $i=1, N_\text{iter}$ do \> \> \> \> {\it main loop (POCS/descent loop)} \\
7:  \>     \> $\vec{f}_0 :=  \vec{f}$ \\
8:  \>     \> for $j = 1, N_d$ do: $\vec{f} :=  \vec{f} +\beta
\vec{M}_j \frac{g_j - \vec{M}_j \cdot \vec{f} }{\vec{M}_j \cdot \vec{M}_j}$ \> \> \> \> {\it ART}\\
9:  \>     \> $\vec{f} := P(\vec{f},0,f_\text{max})$ \> \> \> {\it enforce bounding constraints} \\
10: \>     \> $\vec{f}_\text{res} := \vec{f}$ \\
11: \>     \> $dp := |\vec{f} - \vec{f}_0 | $ \\
12: \>     \> $\vec{f}_0 :=  \vec{f}$ \\
13: \>     \> for $j = 1, ng$ do      \> \> \>{\it steepest descent loop} \\
14: \>     \>     \> $R_0 :=  R( \vec{f}) $ \\
15: \>     \>     \> $\vec{df} := \nabla_{\vec{f}}  R(\vec{f}) $ \\
16: \>     \>     \> $\hat{df} := \vec{df}/|\vec{df}|$ \\
17: \>     \>     \> $\vec{f}^\prime := \vec{f}- dp* \hat{df}$ \\
18: \>     \>     \> $\vec{f}^\prime := P(\vec{f}^\prime,0,f_\text{max})$ \\
19: \>     \>     \> $\gamma := 1.0$ \\
20: \>     \>     \> while  $R(\vec{f}^\prime)>R_0$ do  \> \> {\it projected line search} \\
21: \>     \>     \>       \> $\gamma := \gamma * \gamma_\text{red}$ \\
22: \>     \>     \>       \> $\vec{f}^\prime := \vec{f}- \gamma dp* \hat{df}$ \\
23: \>     \>     \>       \> $\vec{f}^\prime := P(\vec{f}^\prime,0,f_\text{max})$ \\
24: \>     \>     \> end while  \\
25: \>     \>     \> $\vec{f} := \vec{f}^\prime$ \\
26: \>     \> end for  \\
27: \>     \> $dg := |\vec{f}^\prime - \vec{f}_0 | $ \\
28: \>     \> if $dg>r_\text{max} * dp $  then
 $\vec{f} := r_\text{max}\frac{dp}{dg}(\vec{f}^\prime - \vec{f}_0 ) + \vec{f}_0 $ \\
29: \> end for\\
30: \> return $\vec{f}_\text{res}$
\end{tabbing}
\end{tt}
The primary controls of the ASD-POCS algorithm are the parameters $\beta$ and $N_\text{iter}$
on line 1.  As $\beta$ is lowered from a value of 1.0, the image-estimate regularity is
decreased, and as the $N_\text{iter}$ increases the image-estimate data-error is reduced.
In terms of the $R,\delta$ diagram of Fig. \ref{fig:trajSketch}, $\beta$ is a horizontal
control and $N_\text{iter}$ is a vertical control.

For readers interested in the reasoning behind this version of the ASD-POCS algorithm,
the remainder of this section provides a detailed explanation of the
algorithm roughly in the order of the
pseudo-code, starting with line 8:
Reduction of the data error
is accomplished through ART at line 8, and positivity is enforced by the projection at line 9.
For the results below, we do not enforce an image upper bound, $f_\text{max}=\infty$, because
there is little impact.
In general, the size of the image-change
due to POCS, $dp$ in the pseudo-code, is large relative to the progress made
by steepest descent on $R(\cdot)$ especially when we require that the objective
function be reduced with each steepest descent step. Thus, the algorithm is designed
to make as much progress as possible, in terms of maximizing $dg$,
on steepest descent of $R(\cdot)$.  First, multiple
gradient descent steps are taken with the loop starting at line 13. We found that
$ng=5$ loops makes decent progress. Many more loops than that yields diminishing returns.
This parameter is not critical, and we leave it fixed at 5. Second, the projected line search
at lines 19-24 is slightly unusual in that it is designed to maximize the steepest
descent step-size, $dg$, while not increasing the objective function $R(\cdot)$. Thus, the line
search algorithm will in general not find the minimum of $R(\cdot)$ along the image-change
direction $\hat{df}$ as is normally done with line searches. A relatively large line-search-reduction
parameter, $\gamma_\text{red}:=0.8$, is chosen so that, again, $dg$ will be maximal. Furthermore,
the initial guess for the line-search step-size of $dp$, at line 17, is very aggressive.
Choosing $\gamma_\text{red}:=0.8$ is not critical for the results and we leave it fixed,
but it does impact algorithm efficiency. The image estimate resulting from the steepest descent
section will respect positivity because of the projections at lines 18 and 23.

The adaptive element of this algorithm occurs at line 28. The reasoning goes that as long as
the change in the image due to POCS $dp$ is not less than $dg$, each iteration of the outer
loop will make net progress in reducing the data error. In the early iterations, when $dp$ is
large, the steepest descent on $R(\cdot)$ is allowed to take large steps, thereby quickly reducing
the image regularity measure. At later steps $dg$ is constrained to lower values so that data error
is not increased. We include the ratio parameter $r_\text{max}=1.0$ even though it's not
used here. For applications with very high quality data and when it is feasible to take many more
iterations such as a hundred or more, it may be desirable to set $r_\text{max}<1.0$ in order
to make more progress in reducing data error.  If algorithm efficiency is of no concern, then
the reader is referred to our previous ASD-POCS algorithm \cite{Sidky-CCB:08}, where precise control
over the data-error tolerance $\epsilon$ is afforded.  For the present algorithm the tolerance
parameter $\epsilon$ is traded for iteration number, which ends up being the parameter that
controls data error.  In order to control image regularity, normally the steepest-descent step
would be reduced or increased. But, as it is important to maximize $dg$ for efficiency, we instead
control the POCS step-size.  This is effectively controlled by the relaxation parameter $\beta$.
It is set to 1.0 in the pseudo-code, but we vary this control parameter in a range of 0.1 to 1.0,
below. To summarize, the controls of the algorithm are: iteration number, more iterations reduce
data error; and $\beta$, lower $\beta$ reduces $R(\cdot)$.

The final image $\vec{f}_\text{res}$ is considered to be the one after the POCS steps, at line
10, and this is the one shown in the present results. But we point out that there is a non-negligible
difference between this image and the image estimate following the steepest descent \cite{Sidky-TV:06}.
We point out also, that we do not claim this algorithm is optimal in any sense. We regard ASD-POCS
as a framework for generating specific image-reconstruction algorithms.  The adaptive control
step, line 28, can be done differently. For example, in our previous algorithm in Ref. \cite{Sidky-CCB:08}
the data error of the current image estimate is compared against a pre-set data tolerance $\epsilon$.
Also, different convex constraints on the image function can be included in $P$, i.e, different
bounds or support constraints. 

Before going on to the results, we mention a few points about algorithm efficiency.
As written above, the pseudo-code is quite inefficient for the early iterations
of the steepest-descent line-search. At line 20 it is likely that $R(\vec{f}^\prime)>>R_0$,
so it may be desirable to include extra logic that allows much smaller values of $\gamma_\text{red}$
when this is the case, switching back to the larger value when $R(\vec{f}^\prime)$ is
near $R_0$. The pseudo-code above is presented above with simplicity in
mind, so there is no doubt that other such tricks could substantially improve run time.
Computation of the gradient of $R(\vec{f})$ in line 15 is easily implemented on commodity
graphics hardware \cite{Sidky-GPU:07}.

\section{application to DBT projection data}
\label{sec:results}

In this section, we employ the practical ASD-POCS image-reconstruction
algorithm to clinical 
DBT projection data obtained on the
GE-MGH instrument.  In the following,
results of the image
reconstruction are displayed for cases containing microcalcifications and masses.
It will be evident that the ASD-POCS algorithm can have a significant impact on
microcalcification imaging.

\subsection{DBT projection data}
\label{sec:scanner}

\begin{figure}[ht]

\begin{minipage}[b]{0.8\linewidth}
\centering
\centerline{\includegraphics[width=7cm,clip=TRUE]{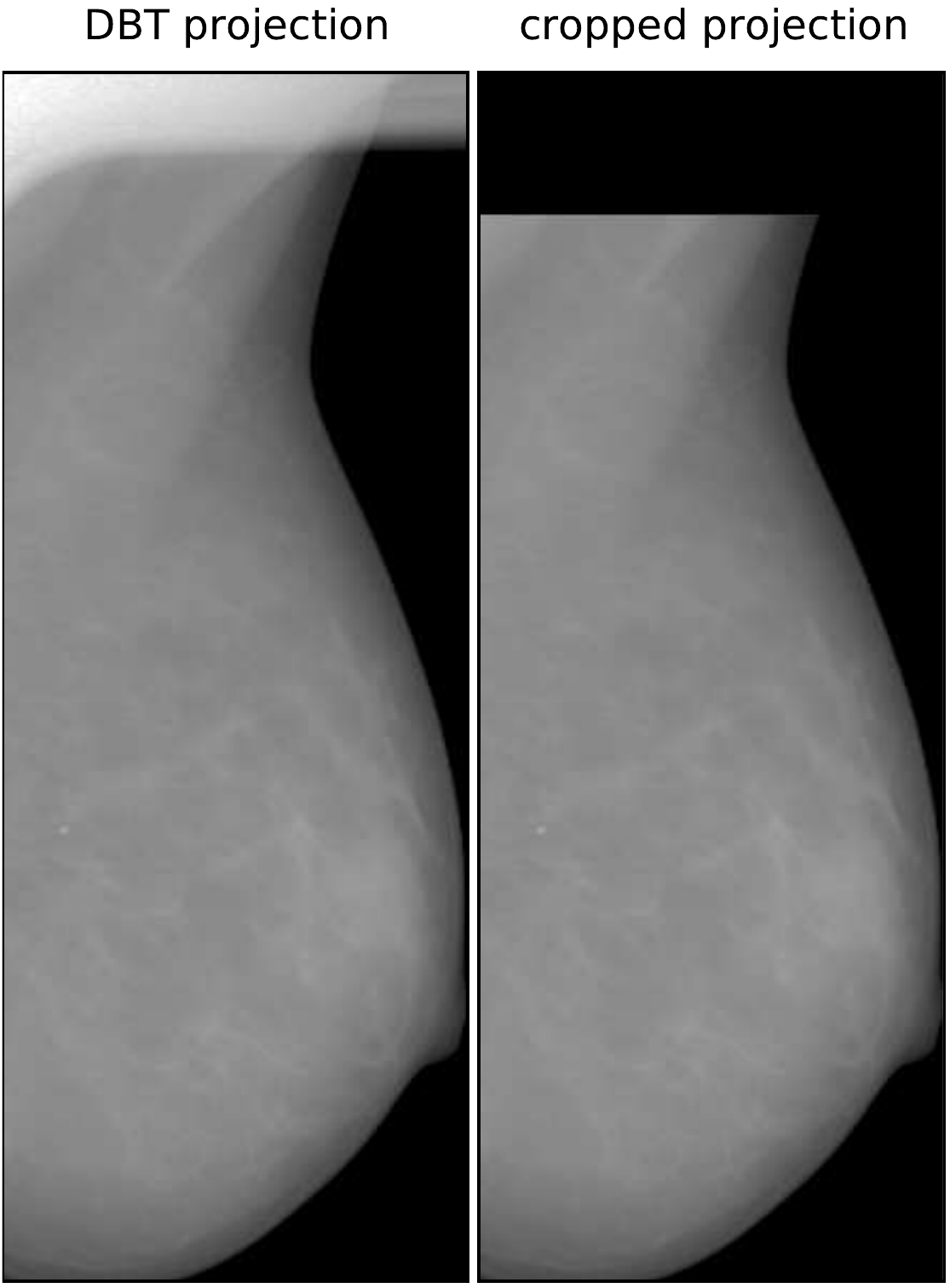}}
\end{minipage}
\caption{(Left) A single projection for the case containing a uniform mass.
(Right) Cropped view used for reconstruction.
\label{fig:projection}}
\end{figure}

As stated earlier, the scan consists of 11 projection views acquired over
a 50$^\circ$ arc. The geometry of the system is shown in Fig. \ref{fig:DBTscanner}.
An example projection from this system is shown in
Fig. \ref{fig:projection} for a view offset at 25$^\circ$.
Note that, for this view, a fin from the compression paddle appears in
the projection.  For such views we truncate the projections to eliminate
rays passing through this fin, because the fin is not in the reconstruction
volume. Doing so reduces artifacts at the edge of the reconstruction volume,
and it allows us to demonstrate convergence properties of the ASD-POCS algorithm.

\subsection{Form of the ASD-POCS objective function and algorithm parameters}
\label{sec:reg}

The ASD-POCS algorithm, presented in Sec. \ref{sec:ASD-POCS}, was shown
with a generic objective function.  For DBT image-reconstruction, here,
we employ a total p-variation (TpV) norm of the image as the objective.
The TpV norm of the image, written in terms of image voxel values $f_{i,j,k}$, is
\begin{equation}
\label{tpv}
\left\| \vec{f} \right\|_{TpV} =
\sum_{i,j,k}  \Delta_{i,j,k}^p,
\end{equation}
where
\begin{equation}
\label{ddef}
\Delta_{i,j,k} = 
\sqrt{ (f_{i,j,k}-f_{i-1,j,k})^2 +(f_{i,j,k}-f_{i,j-1,k})^2 +
(f_{i,j,k}-f_{i,j,k-1})^2 + s }.
\end{equation}
The parameter $s$ is set to $10^{-6}$, here, and it is needed to ensure that the TpV norm
is differentiable with respect to voxel value when $p \le 1.0$.
Because $\Delta_{i,j,k}$ involves a backward difference, the summations in Eq. (\ref{tpv})
start at the second voxel number.
For the images reconstructed below, we take the values of $p$ to be 0.8, 1.0 and 2.0 .
The case of $p=0.8$, results in a non-convex norm, and it may have some advantage
for image-reconstruction from incomplete projection data \cite{chartrand-2007-exact,Sidky-MIC-tpv:07}.
When $p=1.0$, the TpV-norm reduces to the standard TV-norm which is convex, and when $p=2.0$
TpV becomes a quadratic, roughness measure, which is commonly used as a penalty term
for iterative image-reconstruction. It is demonstrated in the results that the value of
$p$ has a significant impact on image quality for DBT.
% In fact, $p$ and $\beta$, which
%controls the strength of the regularization, are the primary controls which can be
%varied to optimize the DBT images for various imaging tasks.

The image array used in the reconstruction consists of 60 slices, 1 mm thick, stacked
parallel to the detector. The in-plane voxel width is 0.1 mm, matching the detector resolution.
The in-plane extent of the slices vary with each case because of breast-size variation
(the volume dimensions are given with each case, below).  The imaging volume is unusual in that
the voxels are 10 times longer in depth than their transverse width.  The limited angular range
of the DBT scan does not readily yield much information on depth variations, hence the thick
slices. Two interesting algorithm aspects that we do not explore here are (1) increasing
depth resolution in the imaging volume and (2) employing spatial differencing for the TpV-norm.
Thinner slices may yield improved depth resolution when used in combination with the TpV-norm
for values $p \le 1.0$.  There are also preliminary indications that using spatial differencing
in Eq. (\ref{ddef}), where the voxel differences in each dimension are divided by the corresponding
voxel length, may improve depth resolution.  We have found that these factors make little difference
for the ASD-POCS algorithm when run in the 10-20 iteration range.   But increasing
depth resolution or employing spatial differencing may yield significantly different images that
solve the optimization problem, Eqs. (\ref{rmin}) and (\ref{constraints}).

For completeness, we provide the expression for the voxel-gradient of the objective
function, Eq. (\ref{tpv}), which is needed for the ASD-POCS algorithm at line 15 of
the pseudo-code.  The $i,j,k^{th}$
component of the gradient is given by:
\begin{align}
\partial \left\| \vec{f} \right\|_{TpV} / \partial f_{i,j,k} =&
 \Delta_{i,j,k}^{p-2} (3 f_{i,j,k} - f_{i-1,j,k} - f_{i,j-1,k} -f_{i,j,k-1})+ \notag \\
& \Delta_{i+1,j,k}^{p-2} (f_{i,j,k}-f_{i+1,j,k})+
 \Delta_{i,j+1,k}^{p-2} (f_{i,j,k}-f_{i,j+1,k})+
 \Delta_{i,j,k+1}^{p-2} (f_{i,j,k}-f_{i,j,k+1}).
\end{align}
Note that this expression applies only to interior voxels. At the edges of the imaging
volume the terms that involve voxels outside the imaging volume should be eliminated.

\subsection{Reconstructed images}
\label{sec:recon}

We demonstrate the ASD-POCS algorithm by investigating image-reconstruction
on three sets of DBT clinical data: one that contains microcalcifications and two cases
that have masses.  For each case, images from a basic EM implementation are also shown.
The EM implementation used is given by the following update equation
\begin{equation}
\label{em}
\vec{f}^{(k+1)}=\vec{f}^{(k)} \cdot
\frac
{
        M^T \cdot
        \left( \tilde{g} / (M \cdot \vec{f}^{(k)})  \right)
}
{
        M^T \cdot \tilde{I}
} ,
\end{equation}
where $\tilde{I}$ is a data vector with every
element set to 1, $k$ is the iteration number,
and the image estimate at
$k=0$ is initialized to 1's in each voxel.
We stress that the EM images are shown only to give a rough idea on the performance of
current algorithms. Furthermore, the goal of this article is not to claim that ASD-POCS
yields ``better'' images, because that is a task dependent issue. Although the
results do seem to indicate a potential advantage for microcalcification imaging.
The aim here, however, is mainly
to demonstrate the image-regularization controls of the 
ASD-POCS algorithm. Using these controls, the images can be optimized for different
tasks in future work.

Each of the three cases, below, are reconstructed in the same way, meaning the
same sets of algorithm parameters are used. The exceptions to this are that
the image volume dimensions and the projection data cropping are slightly different
for each case. For the EM results images are shown at 5,10, and 20 iterations, as
iteration number is really the main control for regularization. For ASD-POCS, the
objection function parameter $p$ is set to 0.8, 1.0, and 2.0; lower values of $p$
tend to sharpen edges. The relaxation factor $\beta$ takes on values of 1.0, 0.5,
and 0.1; smaller $\beta$, in general, allows for ASD-POCS to achieve lower values
of the TpV objective. Images for ASD-POCS are also shown for 5, 10,
and 20 iterations.  As will be seen, there is surprisingly little change in the
reconstructed images for these iteration numbers.  In each of the image sets,
a 2D ROI is displayed that shows either microcalcifications or a mass, depending
on the case.
 
\subsection{Case 1: microcalcifications}
\label{sec:calcResults}

\begin{figure}[ht]

\begin{minipage}[b]{0.8\linewidth}
\centering
%\centerline{\includegraphics[width=7cm,clip=TRUE]{figs/calcSlice_EM.eps}}
\centerline{\includegraphics[width=10cm,clip=TRUE]{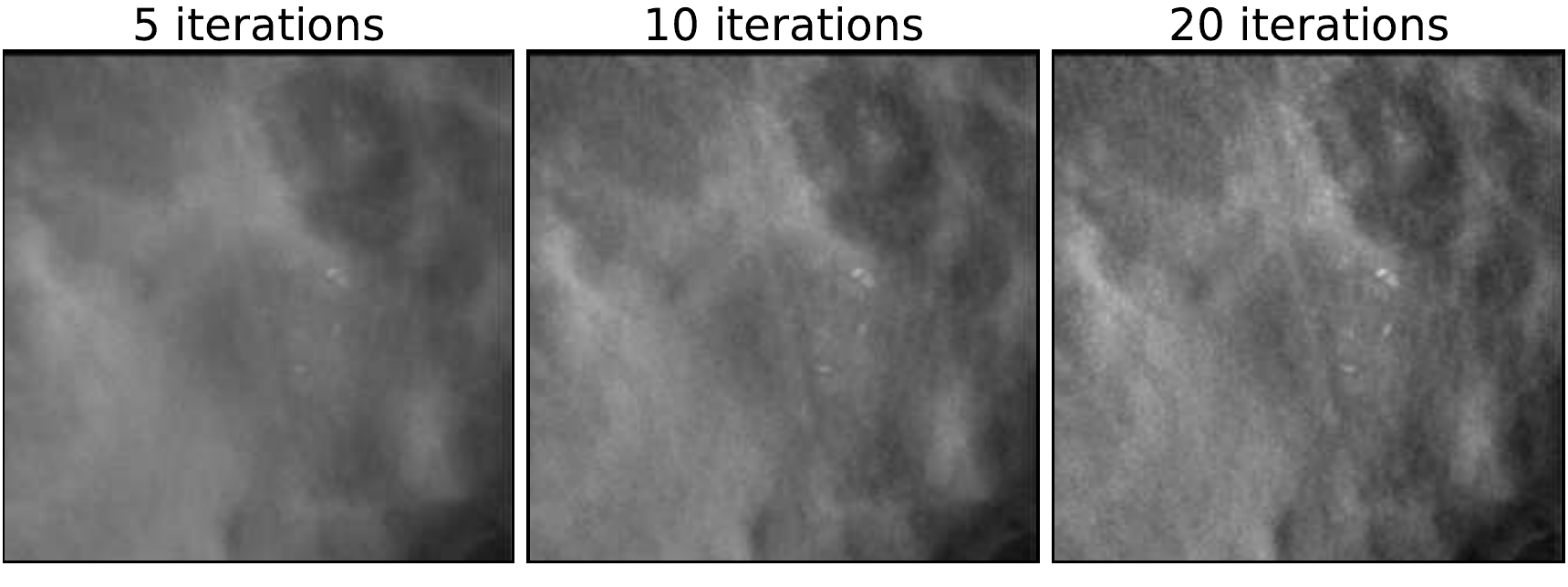}}
\end{minipage}
\caption{
ROI reconstructions of the data set containing microcalcifications
by the EM algorithm at (left) 5, (middle) 10, and (right) 20
iterations. The gray scale window is [0.30,0.65].
\label{fig:calcEM}}
\end{figure}

\begin{figure}[ht]
\begin{minipage}[b]{0.8\linewidth}
\centering
\centerline{\includegraphics[width=10cm,clip=TRUE,angle=90]{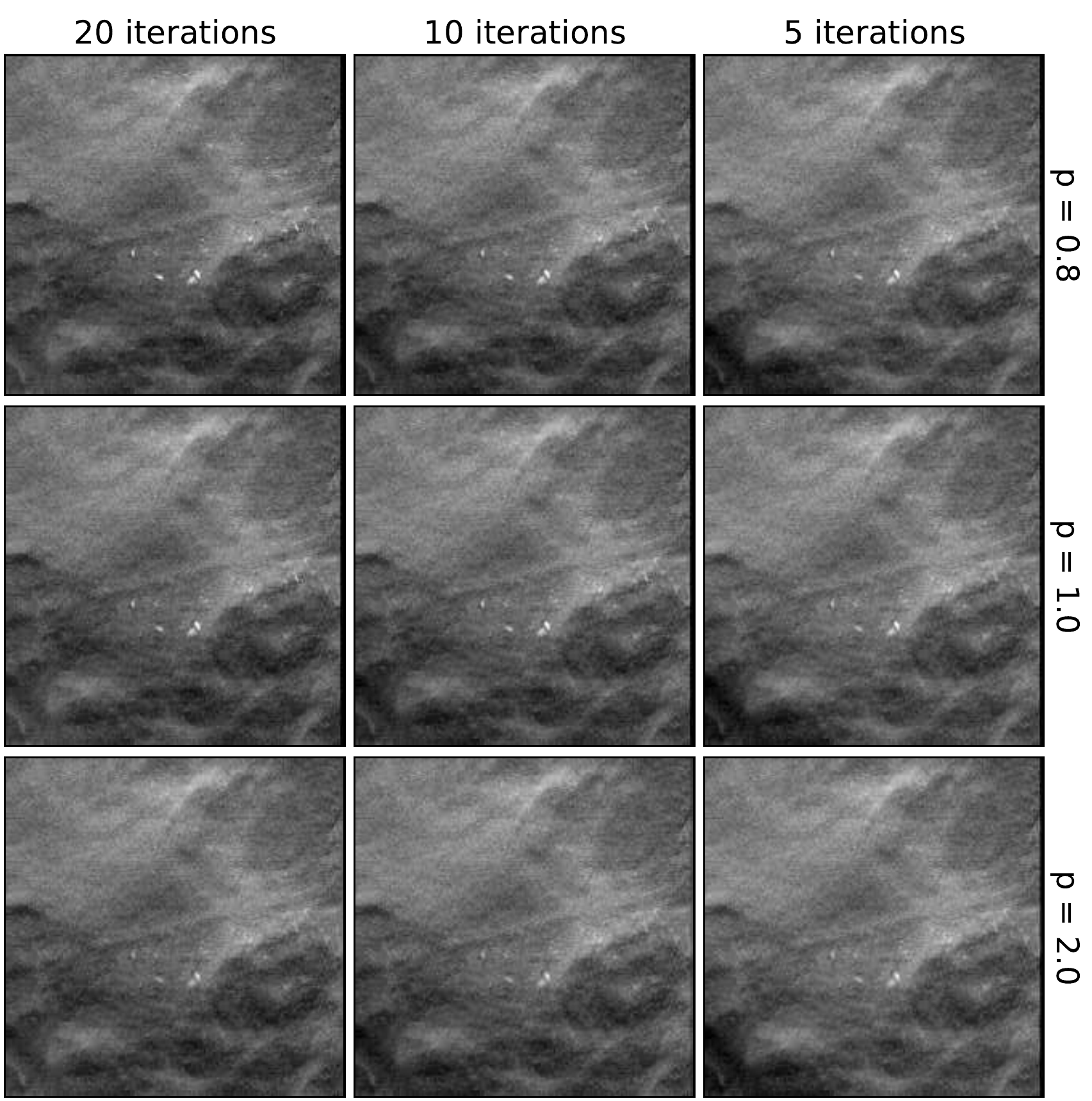}}
\end{minipage}
\caption{
ROI reconstructions of the data set containing a microcalcifications
by the ASD-POCS framework with $\beta = 1.0$.
The gray scale window is held fixed, and is the same as that of the EM results, [0.30,0.65].
\label{fig:calcAPOCS}}
\end{figure}

\begin{figure}[ht]
\begin{minipage}[b]{0.8\linewidth}
\centering
\centerline{\includegraphics[width=10cm,clip=TRUE,angle=90]{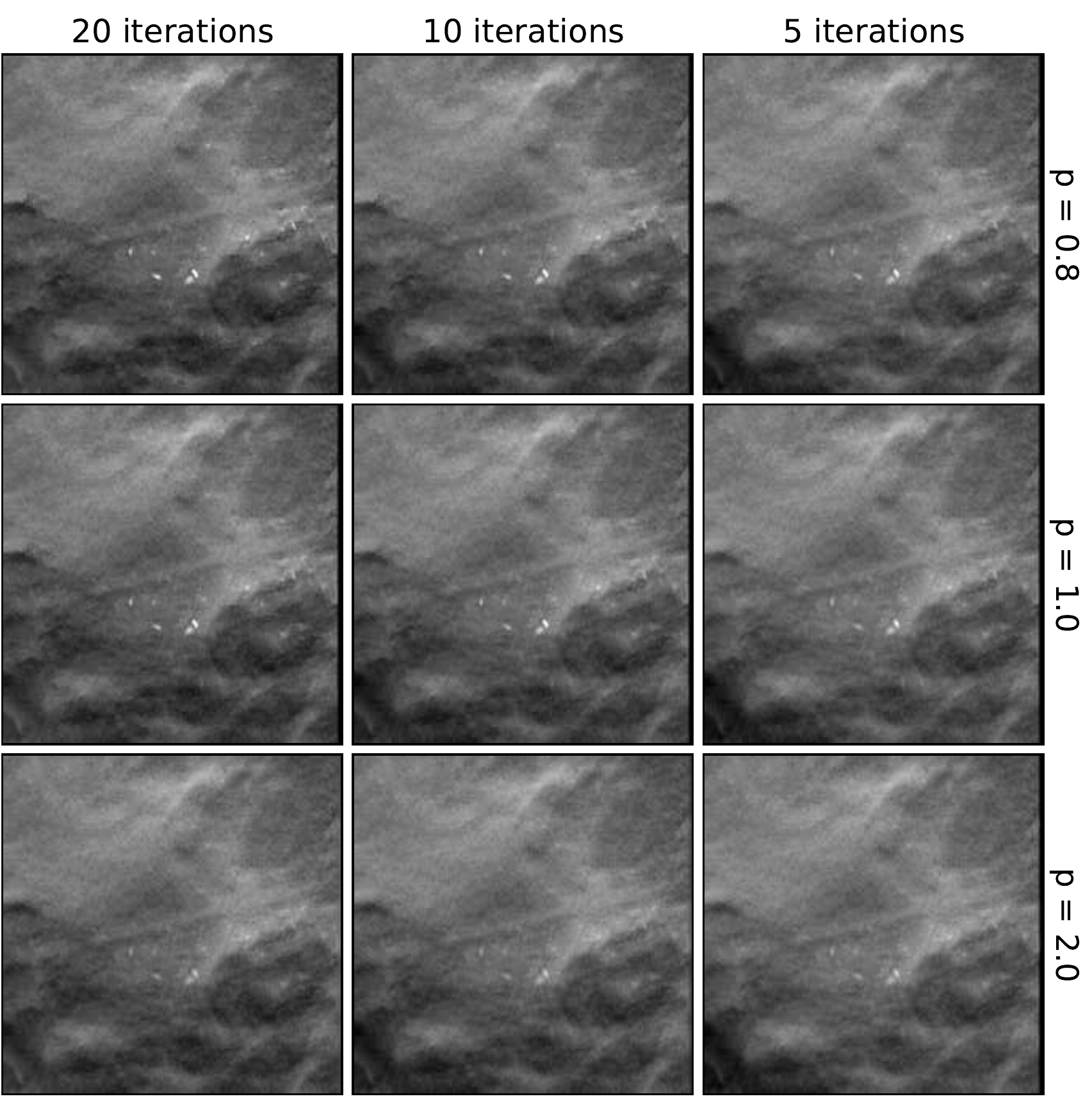}}
\end{minipage}
\caption{
Same as Fig. \ref{fig:calcAPOCS} except $\beta=0.5$.
\label{fig:calcAPOCS1}}
\end{figure}

\begin{figure}[ht]
\begin{minipage}[b]{0.8\linewidth}
\centering
\centerline{\includegraphics[width=10cm,clip=TRUE,angle=90]{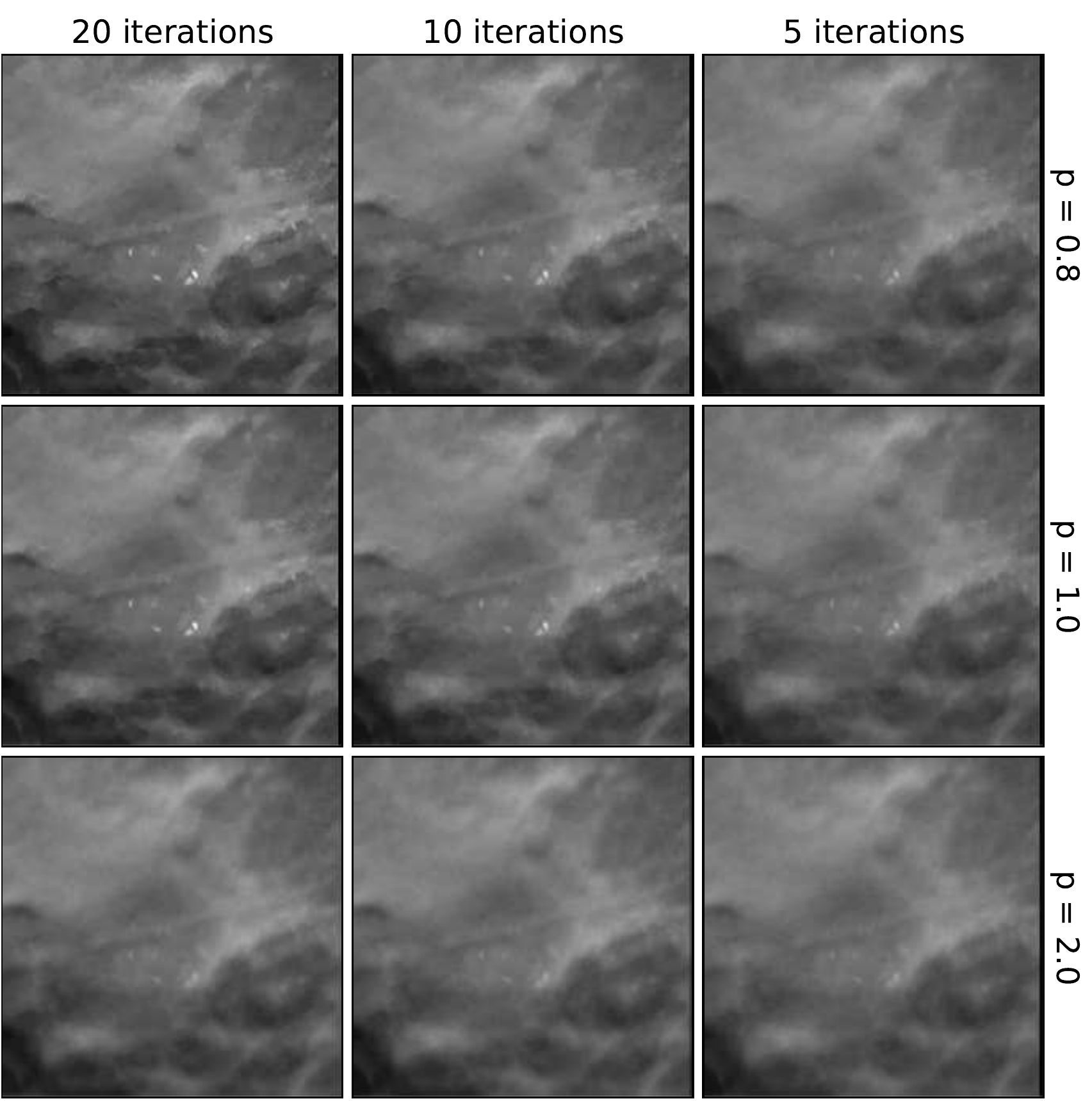}}
\end{minipage}
\caption{
Same as Fig. \ref{fig:calcAPOCS} except $\beta=0.1$.
\label{fig:calcAPOCS2}}
\end{figure}

\begin{figure}[ht]
%\begin{minipage}[b]{0.49\linewidth}
%\centering
%\centerline{\includegraphics[width=8cm,clip=TRUE]{figs/calcTransProf.eps}}
%\end{minipage}
%\begin{minipage}[b]{0.49\linewidth}
%\centering
%\centerline{\includegraphics[width=8cm,clip=TRUE]{figs/calcDepthProf.eps}}
%\end{minipage}
\begin{minipage}[b]{0.9\linewidth}
\centering
\centerline{\includegraphics[width=13cm,clip=TRUE]{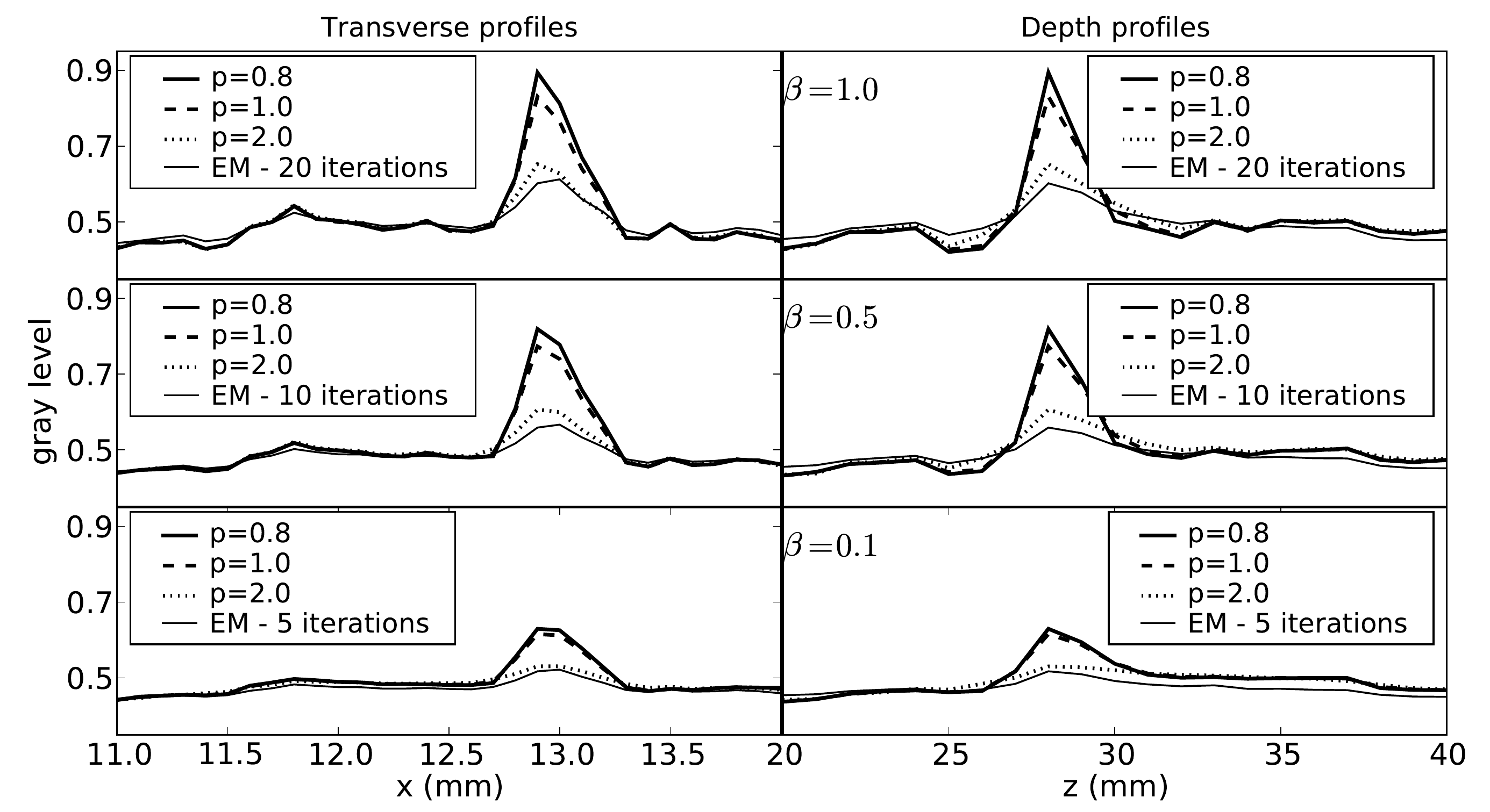}}
\end{minipage}
\caption{Profiles, centered on a microcalcification, through reconstructed images
for different values of $p$ and $\beta$. Also shown are results by the EM algorithm.
The comparison of EM at different iteration number does not necessarily have
any relation to the ASD-POCS results at different $\beta$.
(Top) Transverse profiles along the $x$-direction.
(Bottom) Depth profiles in the $z$-direction.  The fact that microcalcification
have a greater width in the depth profiles is likely to inherent blurring
in the DBT system.
\label{fig:calcProfiles}}
\end{figure}

A set of EM images for the first case is shown in Fig. \ref{fig:calcEM}, and
the corresponding ASD-POCS images are shown in Figs. \ref{fig:calcAPOCS}, \ref{fig:calcAPOCS1},
and \ref{fig:calcAPOCS2}.
A striking feature of the ASD-POCS reconstructions is the prominence of
the microcalcifications.  Lower values of $p$ accentuate these small features
better than large $p$-values.  Even for $p=2.0$, the visibility of the
microcalcifications is comparable to that of the EM results.  The differences
in microcalcification contrast can be seen quantitatively in the 
profiles shown in Fig. \ref{fig:calcProfiles}.  These profiles are plotted
along depth and transverse lines that intersect with a single microcalcification.
We point out that while lower $\beta$ increases regularization strength in ASD-POCS and
lower iteration number increases regularization strength for EM, there is
no direct correspondence between the two parameters; the chosen iteration
numbers for the EM profiles are selected only for reference. Interestingly, there
seems to be little change in the ASD-POCS image for iteration numbers 5-20, which
obviously has some practical implication.

%\begin{figure}[ht]

%\begin{minipage}[b]{0.48\linewidth}
%\centering
%\centerline{\includegraphics[width=7cm,clip=TRUE]{figs/calcSlice_rf1.eps}}
%\end{minipage}
%\begin{minipage}[b]{0.48\linewidth}
%\centering
%\centerline{\includegraphics[width=7cm,clip=TRUE]{figs/calcSlice_rf2.eps}}
%\end{minipage}
%\begin{minipage}[b]{0.48\linewidth}
%\centering
%\centerline{\includegraphics[width=7cm,clip=TRUE]{figs/calcSlice_rf10.eps}}
%\end{minipage}
%\caption{
%ROI reconstructions of the data set containing a uniform mass
%by the ASD-POCS framework. The left, right, and bottom
%panels correspond to $\beta = 1.0$, $\beta = 0.5$, and $\beta = 0.1$, respectively.
%The gray scale window is held fixed, and is the same as that of the EM results, [0.30,0.65].
%\label{fig:calcAPOCS}}
%\end{figure}

From the profiles and slice images, it is clear that lower $p$ in ASD-POCS
enhances microcalcification contrast substantially, leaving one to wonder
if there is any advantage to larger $p$-values.  While lower $p$-values appear
to be advantageous, there is also an impact of $p$-value on the image background.
The ROIs displayed in Figs. \ref{fig:calcAPOCS}, \ref{fig:calcAPOCS1}, and \ref{fig:calcAPOCS2}
are shown in a large enough region
to obtain some sense of the difference in background. Again, we are trying, here,
to only give some intuition on the parameter-space $(p,\beta)$
dependence of the ASD-POCS algorithm.  Optimal values of $p$ and $\beta$ for particular
tasks, such as microcalcification detection by human observers, need to be investigated
in separate studies.  Another important factor that affects selection of $p$ and $\beta$
is data quality. Lower values of $p$, for example, may be robust against detector noise,
but may be also more sensitive to inconsistency due to patient motion.  

If, upon further study, it turns out that low $p$ image-reconstruction with ASD-POCS
consistently yields improved contrast on microcalcification imaging, the implication
for DBT imaging is enormous.  It is known that microcalcification imaging is noise-limited,
while mass imaging is structured-background limited.  Image reconstruction algorithms
that increase microcalcification detectability {\it may lower the required intensity of the
probing X-ray beam}, thus lowering the radiation-dose of the DBT scan.

\subsection{Case 2: uniform mass}
\label{sec:mass2Results}

\begin{figure}[ht]

\begin{minipage}[b]{0.48\linewidth}
\centering
\centerline{\includegraphics[width=10cm,clip=TRUE]{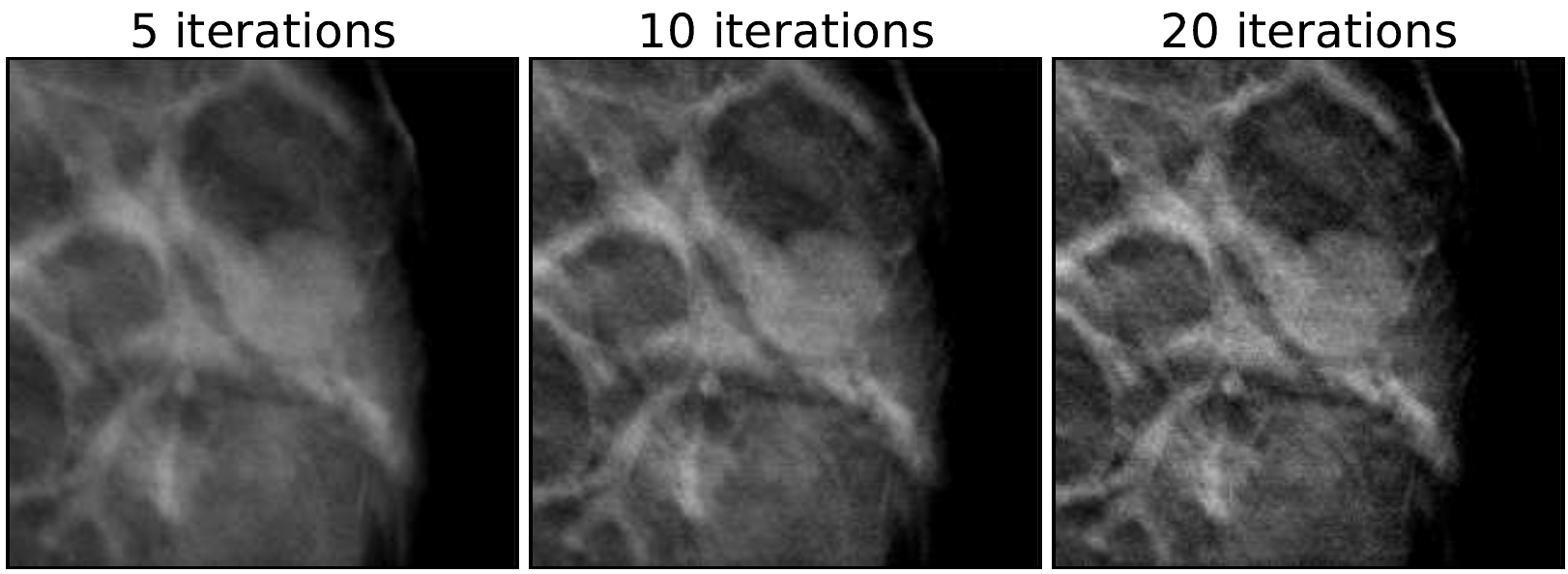}}
\end{minipage}
\caption{
ROI reconstructions of the data set containing a uniform mass
by the EM algorithm at (left) 5, (middle) 10, and (right) 20
iterations. The gray scale window is [0.35,0.55].
\label{fig:mass1EM}}
\end{figure}

\begin{figure}[ht]

\begin{minipage}[b]{0.8\linewidth}
\centering
\centerline{\includegraphics[width=10cm,clip=true,angle=90]{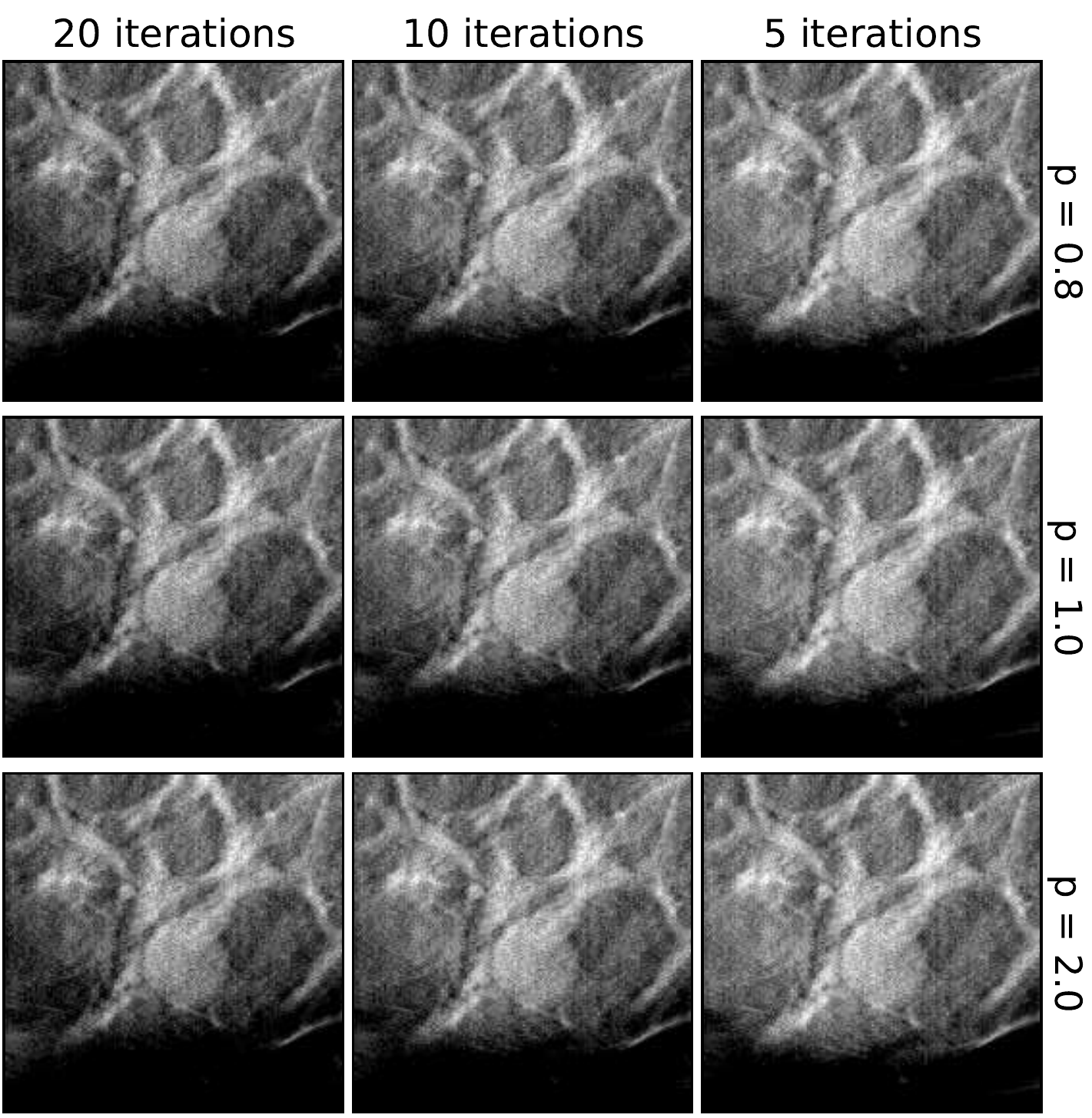}}
\end{minipage}
\caption{
ROI reconstructions of the data set containing the uniform mass
by the ASD-POCS framework with $\beta = 1.0$.
The gray scale window is held fixed, and is the same as that of the EM results, [0.35,0.55].
\label{fig:mass1APOCS}}
\end{figure}

\begin{figure}[ht]

\begin{minipage}[b]{0.8\linewidth}
\centering
\centerline{\includegraphics[width=10cm,clip=true,angle=90]{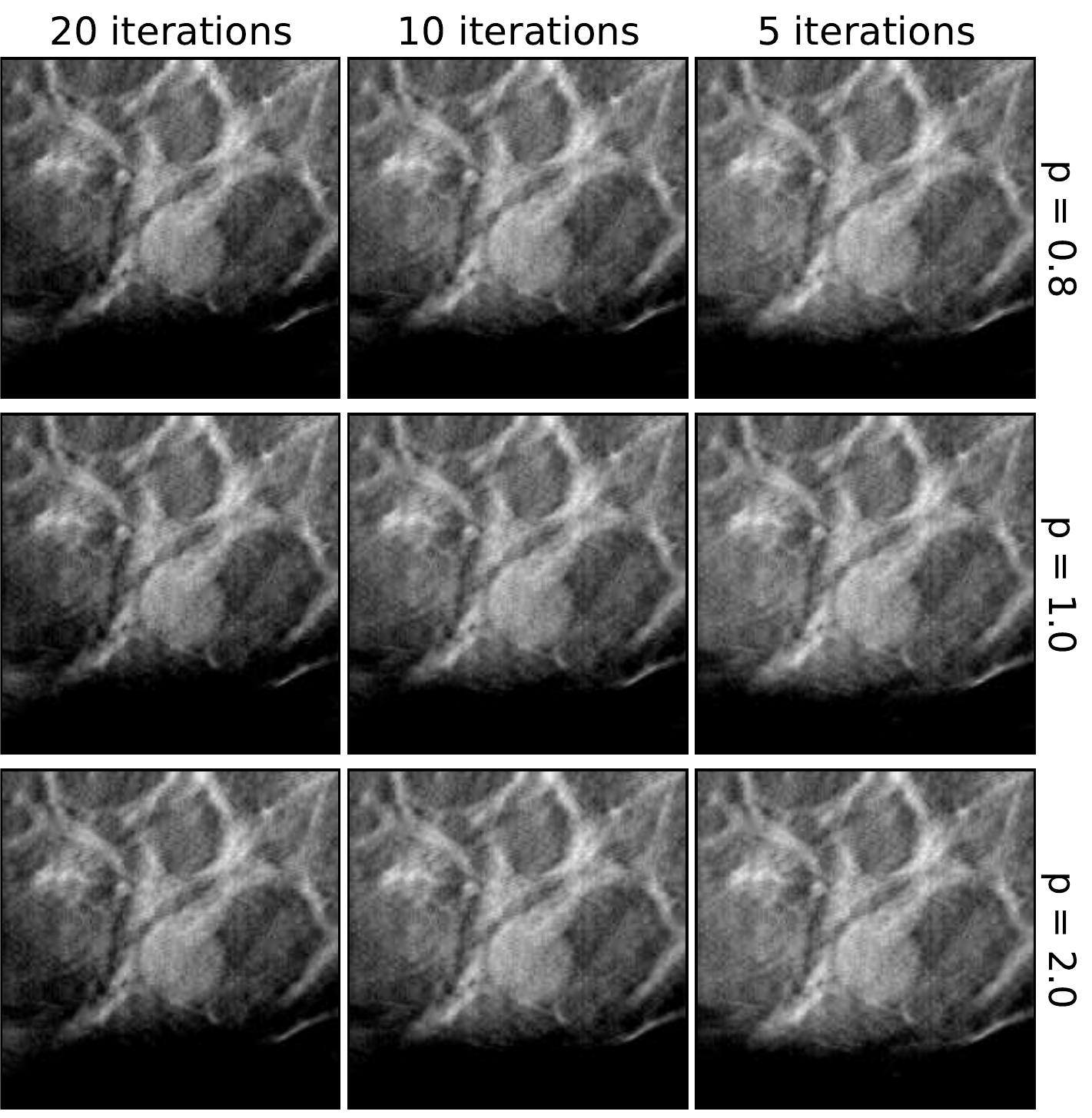}}
\end{minipage}
\caption{
Same as Fig. \ref{fig:mass1APOCS} except $\beta = 0.5$.
\label{fig:mass1APOCS1}}
\end{figure}

\begin{figure}[ht]

\begin{minipage}[b]{0.8\linewidth}
\centering
\centerline{\includegraphics[width=10cm,clip=true,angle=90]{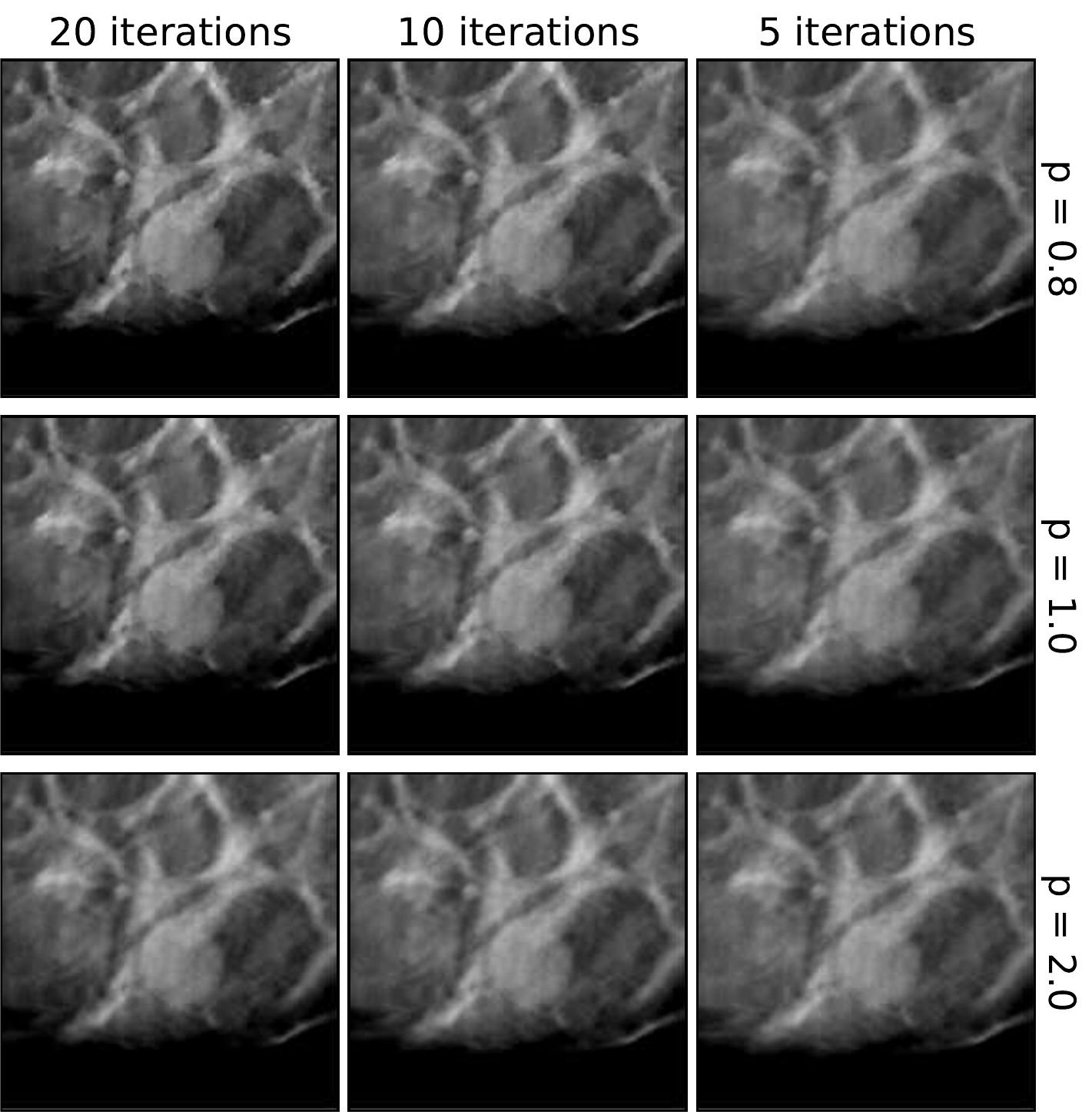}}
\end{minipage}
\caption{
Same as Fig. \ref{fig:mass1APOCS} except $\beta = 0.1$.
\label{fig:mass1APOCS2}}
\end{figure}

For the next case, there is a uniform mass, as can be seen in the EM image-reconstructions
in Fig. \ref{fig:mass1EM}.  As was done in the previous case, we present a spread of images
in Figs. \ref{fig:mass1APOCS}, \ref{fig:mass1APOCS1}, and \ref{fig:mass1APOCS2}
from the ASD-POCS algorithm for the same sets of algorithm parameters, covering a range
of $p$- and $\beta$-values.
The iteration number
dependence appears to be weak for ASD-POCS.  
The conspicuity of the mass for this case does not vary with
algorithm parameters nearly as much as the microcalcification conspicuity of the previous
case.  There are many reasons for this. First, the X-ray attenuation coefficient of the mass
is less than that of calcium, so the contrast that can be potentially regained is not as great.
Second, the lower $p$ reconstructions tend to yield sharper edges, but this does not have
as large an effect on the mass which is substantially bigger than microcalcifications.
Finally, as pointed out earlier, mass conspicuity tends to depend on background structure noise.
As this type of background is physically there, low $p$ image-reconstruction sharpens
the edges of the background features just as much as the mass's edges. Thus, the conspicuity
of the mass may not improve dramatically as $p$ is lowered. In any case, there are subtle
differences between the images, and these differences may have an impact on human or machine
observers.

%\begin{figure}[ht]

%\begin{minipage}[b]{0.48\linewidth}
%\centering
%\centerline{\includegraphics[width=7cm,clip=true]{figs/mass2slice_rf1.eps}}
%\end{minipage}
%\begin{minipage}[b]{0.48\linewidth}
%\centering
%\centerline{\includegraphics[width=7cm,clip=true]{figs/mass2slice_rf2.eps}}
%\end{minipage}
%\begin{minipage}[b]{0.48\linewidth}
%\centering
%\centerline{\includegraphics[width=7cm,clip=true]{figs/mass2slice_rf10.eps}}
%\end{minipage}
%\caption{
%roi reconstructions of the data set containing the uniform mass
%by the asd-pocs framework. the top, middle, and bottom
%panels correspond to $\beta = 1.0$, $\beta = 0.5$, and $\beta = 0.1$, respectively.
%the gray scale window is held fixed, and is the same as that of the em results, [0.35,0.55].
%\label{fig:mass1apocs}}
%\end{figure}

Comparing the visual quality of the images of the present case with the previous one, it
is interesting that similar $\beta$-values do not yield similar apparent image quality.
For example, $\beta=1.0$ for the present case appears to be quite noisy, even taking into
account differing gray level windows, relative to $\beta=1.0$ for the previous case.  For
the 3 sets of $\beta$-values, $\beta=0.1$ appears to yield, visually, the best images for
this mass case, while $\beta=0.5$ seems to best for the previous, microcalcification case.
These, differences are likely due to varying quality of the acquired projection data. 
A quantitative discussion of algorithm performance across different DBT cases will be
further elaborated on in Sec. \ref{sec:evol}.

\subsection{Case 3: spiculated mass in a dense breast}
\label{sec:mass1Results}

\begin{figure}[ht]

\begin{minipage}[b]{0.8\linewidth}
\centering
\centerline{\includegraphics[width=10cm,clip=TRUE]{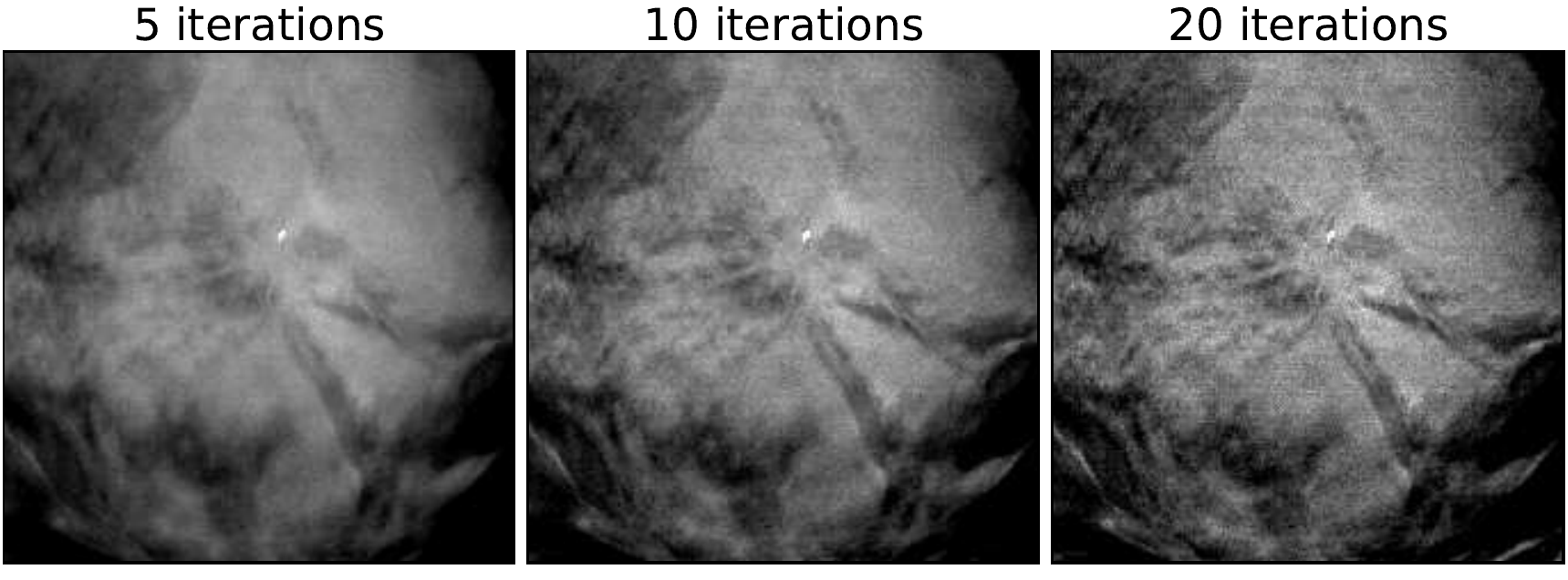}}
\end{minipage}
\caption{
ROI reconstructions of the data set containing a spiculated mass in a dense breast
by the EM algorithm at (left) 5, (middle) 10, and (right) 20
iterations. The gray scale window is [0.42,0.57].
\label{fig:mass2EM}}
\end{figure}

\begin{figure}[ht]

\begin{minipage}[b]{0.8\linewidth}
\centering
\centerline{\includegraphics[width=10cm,clip=TRUE,angle=90]{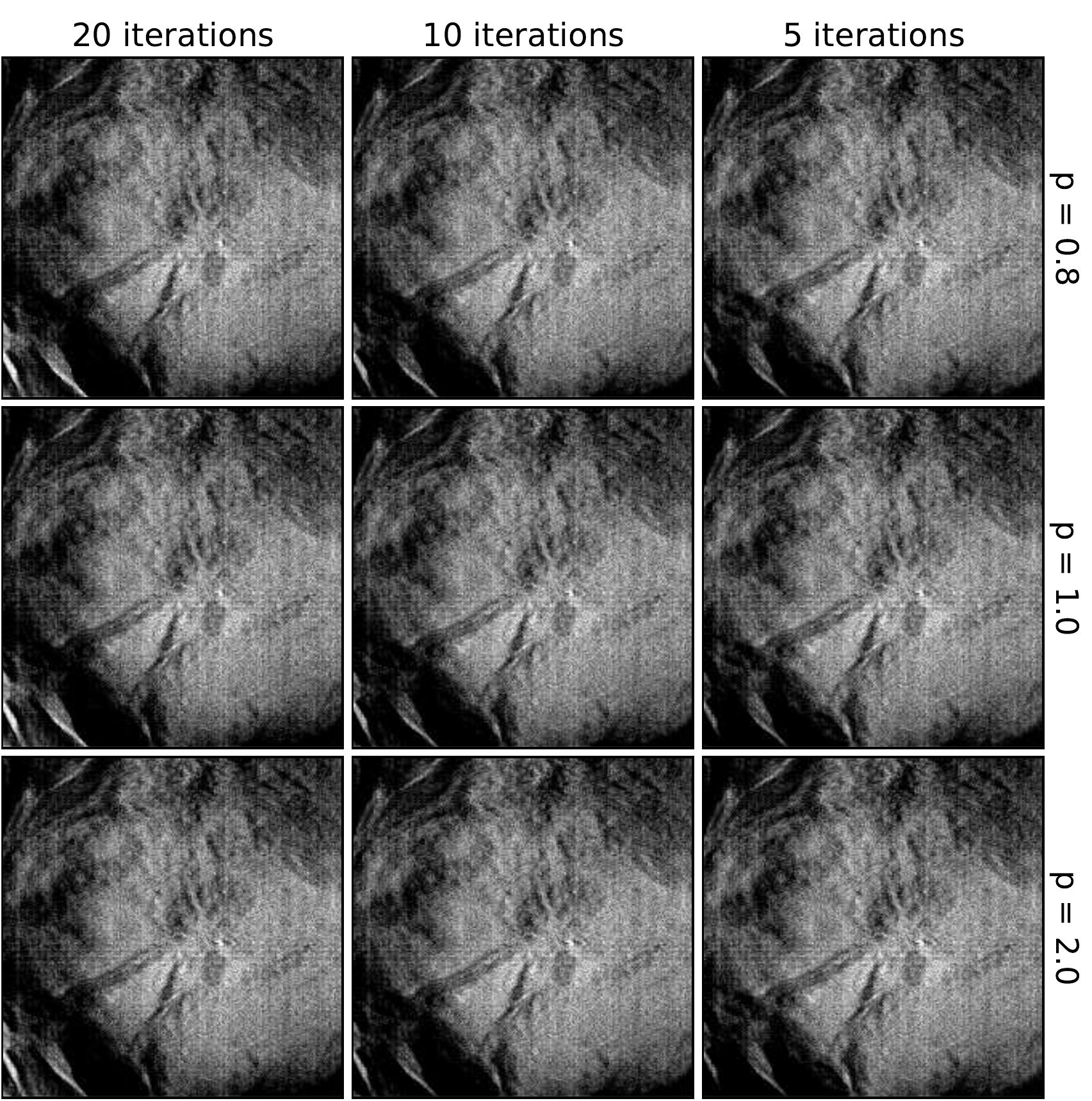}}
\end{minipage}
\caption{
ROI reconstructions of the data set containing the spiculated mass in a dense breast
by the ASD-POCS framework with $\beta = 1.0$.
The gray scale window is held fixed, and is the same as that of the EM results, [0.42,0.57].
\label{fig:mass2APOCS}}
\end{figure}

\begin{figure}[ht]

\begin{minipage}[b]{0.8\linewidth}
\centering
\centerline{\includegraphics[width=10cm,clip=TRUE,angle=90]{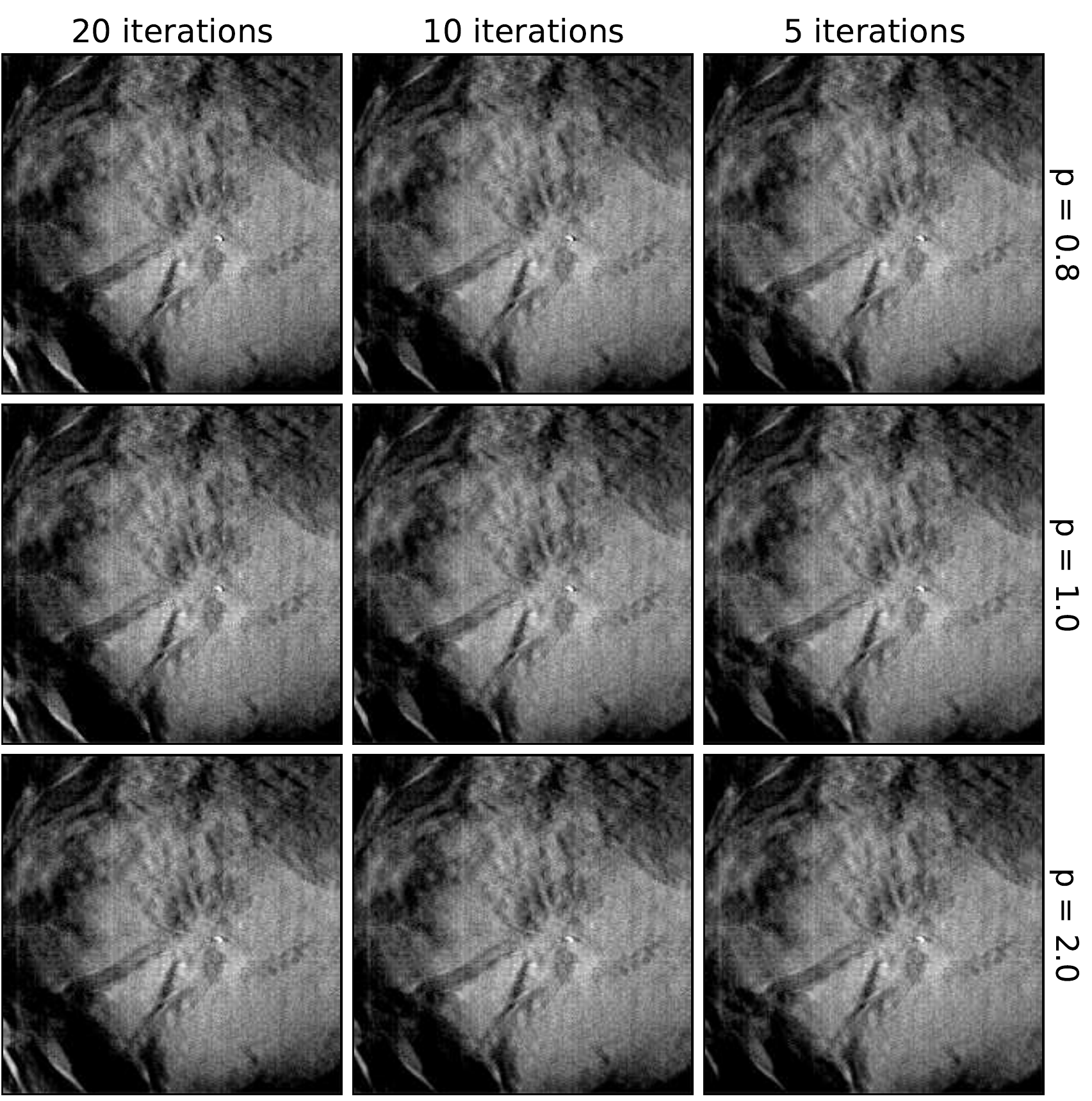}}
\end{minipage}
\caption{
Same as Fig. \ref{fig:mass2APOCS} except $\beta = 0.5$.
\label{fig:mass2APOCS1}}
\end{figure}

\begin{figure}[ht]

\begin{minipage}[b]{0.8\linewidth}
\centering
\centerline{\includegraphics[width=10cm,clip=TRUE,angle=90]{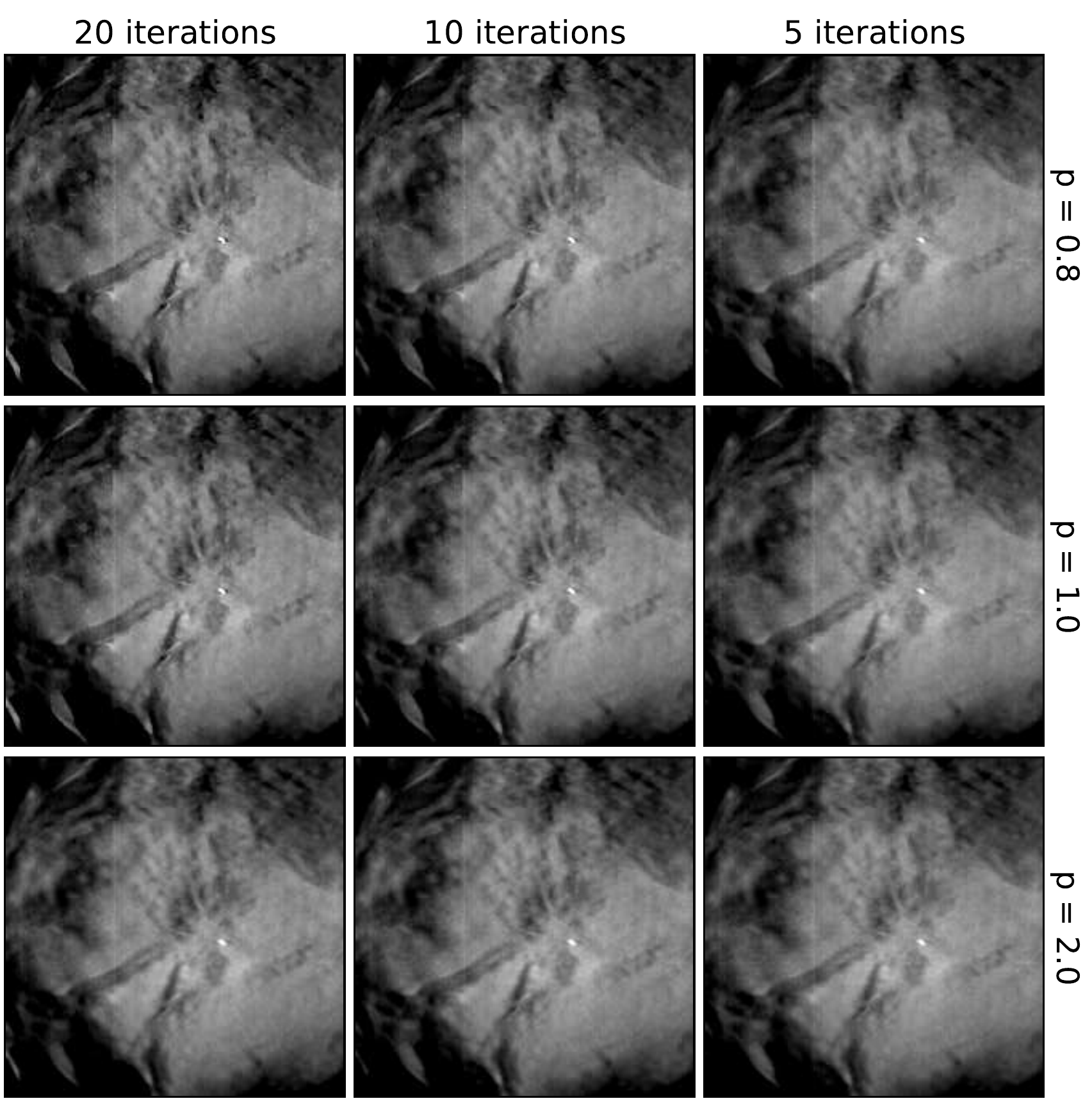}}
\end{minipage}
\caption{
Same as Fig. \ref{fig:mass2APOCS} except $\beta = 0.1$.
\label{fig:mass2APOCS2}}
\end{figure}

Finally, we present a case with a spiculated mass in dense breast tissue. It is
precisely this type of case which DBT was developed for; by removing some of the
interference of the overlapping structures such masses may be more conspicuous
in DBT images than in standard mammographic projection imaging.  The EM images
are shown in Fig. \ref{fig:mass2EM}, and the ASD-POCS images are shown in
Figs. \ref{fig:mass2APOCS}, \ref{fig:mass2APOCS1}, and \ref{fig:mass2APOCS2}.
As with the previous mass case, there may be some
advantage to image-reconstruction with ASD-POCS at low $p$ due to the fact that
edges are enhanced. But the advantage is not as clear cut as it is with
microcalcification imaging.  Any advantage in mass imaging needs to be demonstrated
by task-based image quality evaluation.

%\begin{figure}[ht]

%\begin{minipage}[b]{0.48\linewidth}
%\centering
%\centerline{\includegraphics[width=7cm,clip=TRUE]{figs/massSlice_rf1.eps}}
%\end{minipage}
%\begin{minipage}[b]{0.48\linewidth}
%\centering
%\centerline{\includegraphics[width=7cm,clip=TRUE]{figs/massSlice_rf2.eps}}
%\end{minipage}
%\begin{minipage}[b]{0.48\linewidth}
%\centering
%\centerline{\includegraphics[width=7cm,clip=TRUE]{figs/massSlice_rf10.eps}}
%\end{minipage}
%\caption{
%ROI reconstructions of the data set containing the spiculated mass in a dense breast
%by the ASD-POCS framework. The top, middle, and bottom
%panels correspond to $\beta = 1.0$, $\beta = 0.5$, and $\beta = 0.1$, respectively.
%The gray scale window is held fixed, and is the same as that of the EM results, [0.42,0.57].
%\label{fig:mass2APOCS}}
%\end{figure}

With this case, under-regularization, at large $\beta$, tends to yield linear artifacts
in the image.  Actually, similar lines appear for the other cases in the first two
iterations of ASD-POCS, but the quickly disappear and are gone by the fifth iteration.
These lines, for the present case, are likely due to a slight system misalignment or patient motion.
This case reveals the control afforded by the $\beta$ parameter in the ASD-POCS
algorithm.  It is easy to select a value of $\beta$ small enough to wash out the
linear artifacts without severely blurring the underlying features of the image.

\subsection{Evolution of algorithm metrics}
\label{sec:evol}

\begin{figure}[ht]

\begin{minipage}[b]{0.8\linewidth}
\centering
%\centerline{\includegraphics[width=6cm,clip=TRUE]{figs/calctraj.eps}}
\centerline{\includegraphics[width=8cm,clip=TRUE]{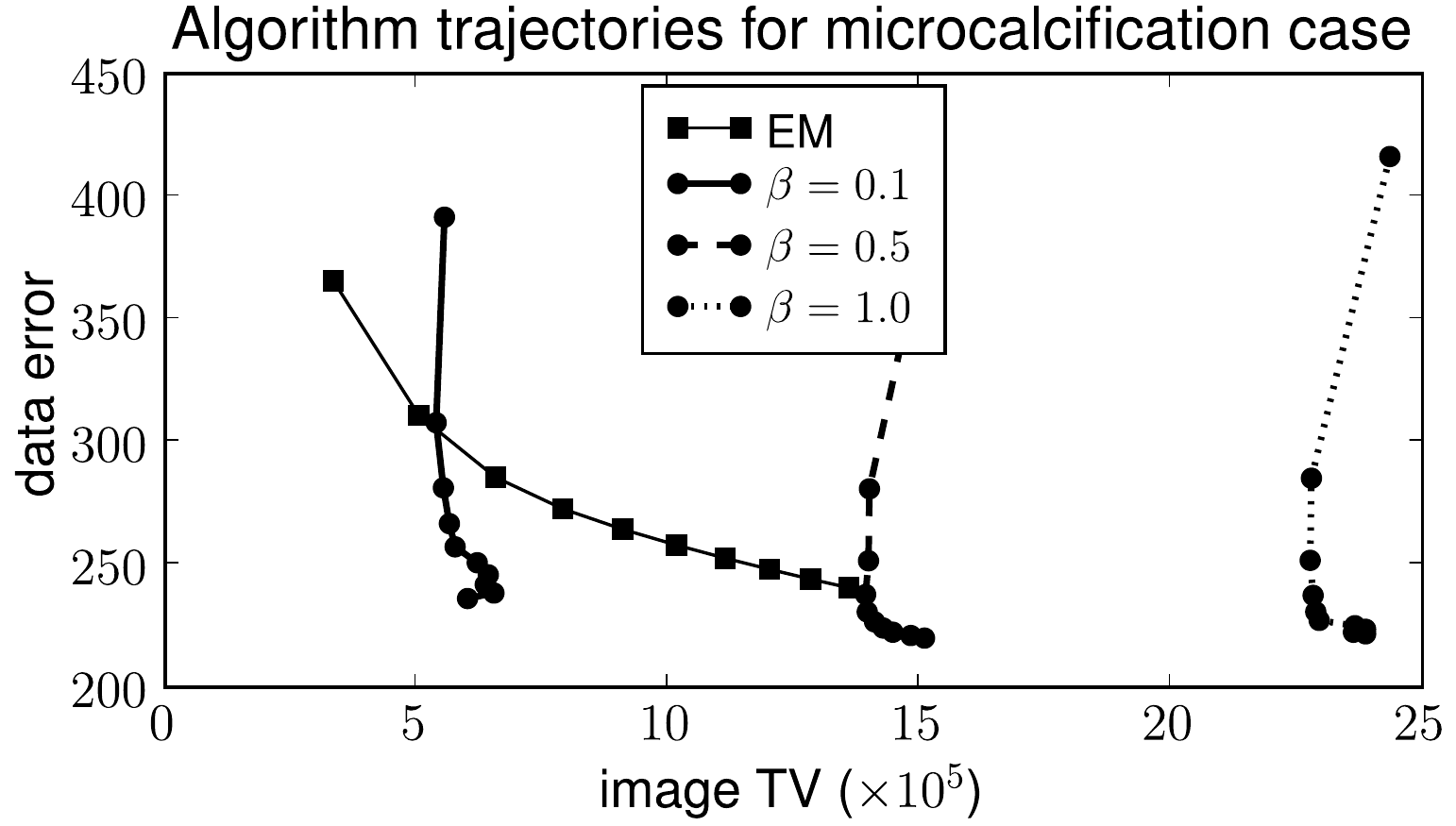}}
\end{minipage}
\vskip 0.2in
\begin{minipage}[b]{0.8\linewidth}
\centering
%\centerline{\includegraphics[width=6cm,clip=TRUE]{figs/mass2traj.eps}}
\centerline{\includegraphics[width=8cm,clip=TRUE]{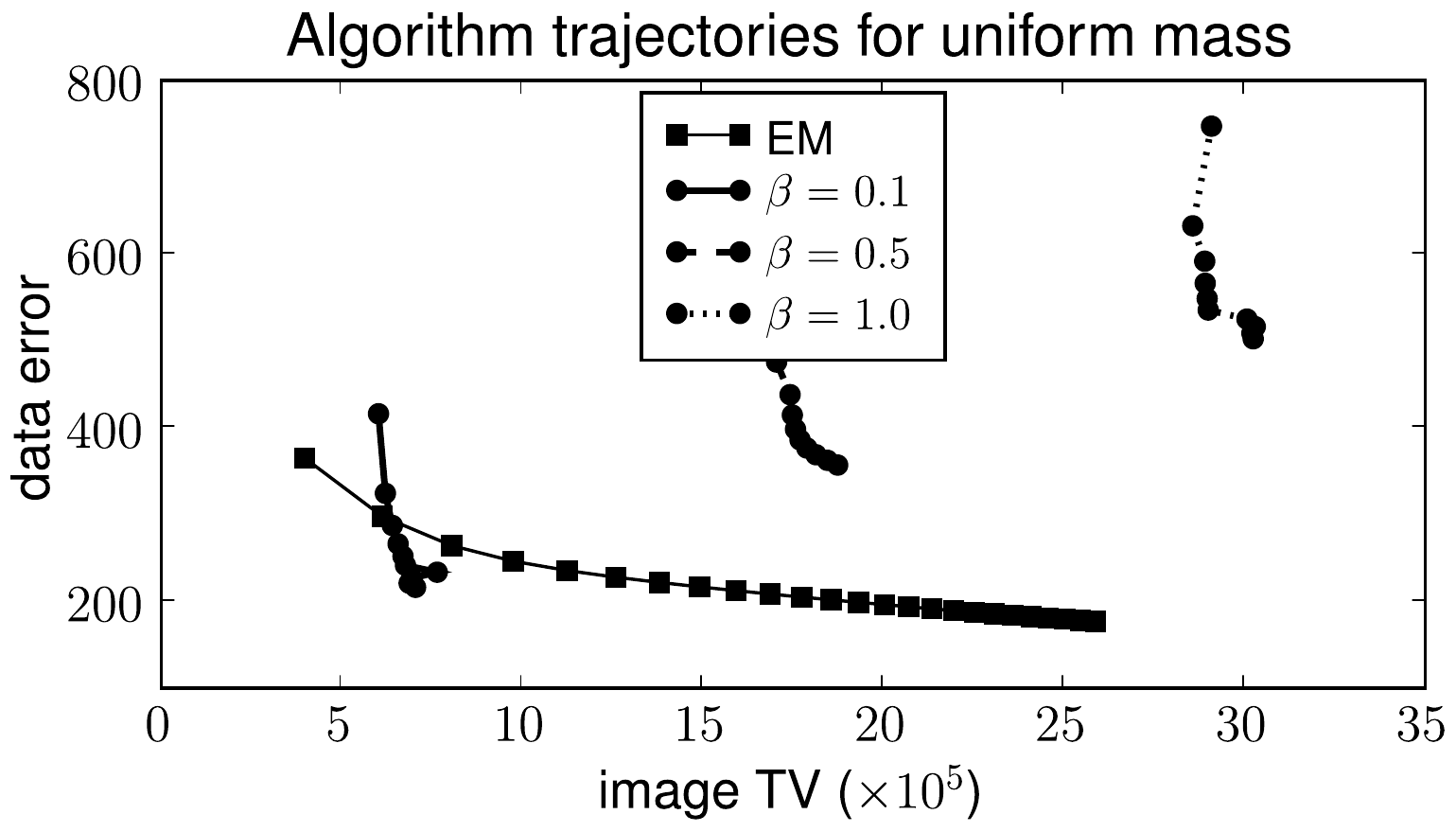}}
\end{minipage}
\vskip 0.2in
\begin{minipage}[b]{0.8\linewidth}
\centering
%\centerline{\includegraphics[width=6cm,clip=TRUE]{figs/masstraj.eps}}
\centerline{\includegraphics[width=8cm,clip=TRUE]{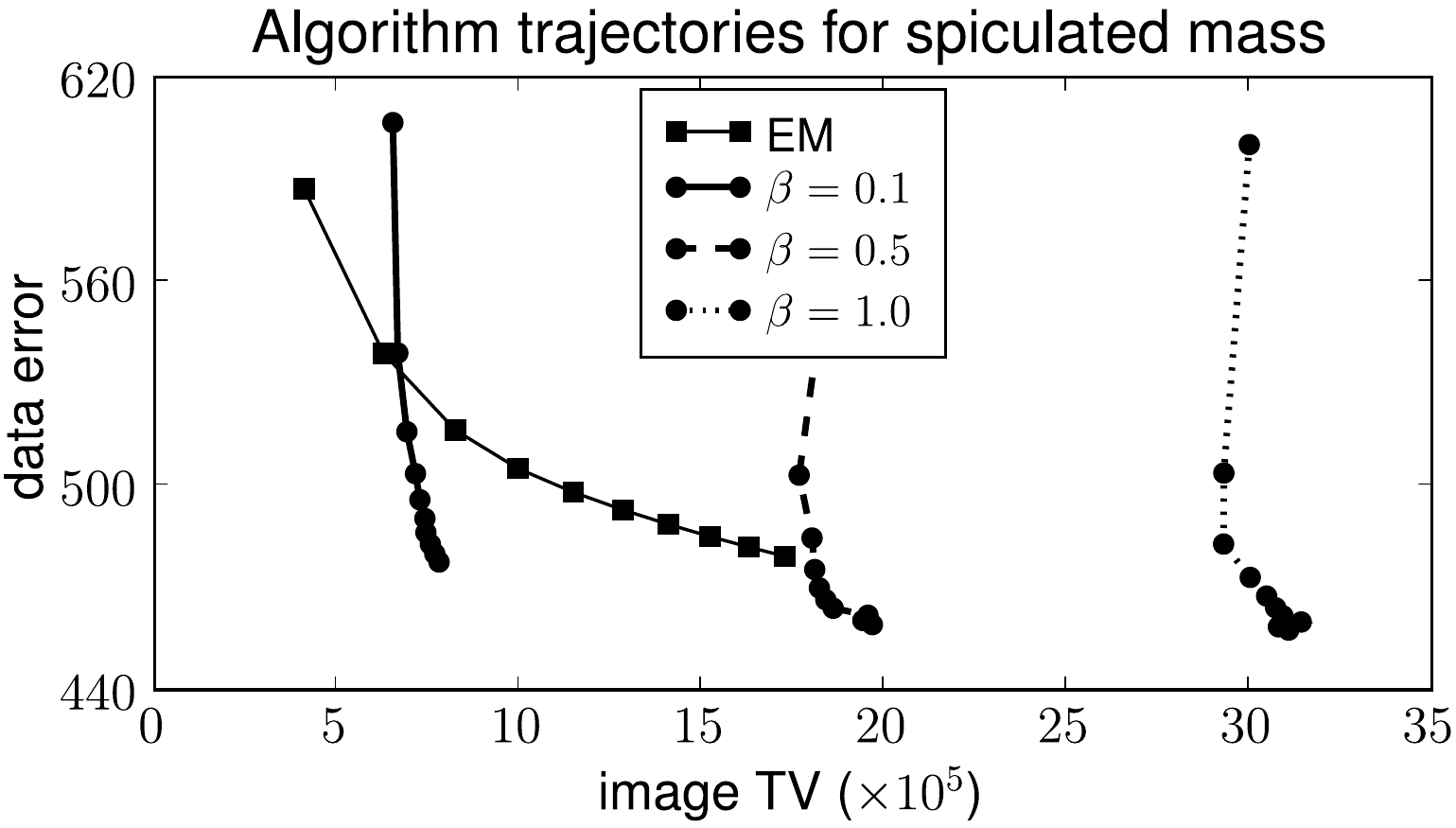}}
\end{minipage}
\caption{
ASD-POCS versus EM parameter trajectories for the data sets containing
(top) microcalcifications, (middle) a uniform mass, and (bottom) a spiculated
mass. ASD-POCS with only $p=1.0$ is shown.  In each case, the actual
iteration numbers are indicated by the symbols starting at iteration 2, at the top
of each curve, and increasing by 2 until 20 iterations at the bottom of
the curves.
\label{fig:traj}}
\end{figure}

It is instructive to return to the discussion on the ASD-POCS algorithm, and
examine the trajectories of the image estimates in the $R,\delta$-plane.
Figure \ref{fig:traj} shows this evolution for each of the three DBT cases
for $p=1.0$. 
The plotted data error is given by:
\begin{equation}
\label{dataerror}
\delta = \sqrt{ (M \vec{f} - \tilde{g}) \cdot (M \vec{f} - \tilde{g})},
\end{equation}
and the objective function $R(\cdot)$ is Eq. (\ref{tpv}) with $p=1.0$.
It is primarily for the
purpose of generating these graphs that the projection rays intersecting the compression
paddle were excluded from the DBT projection data sets.  Retaining these inconsistent
rays would skew the values of the data error.  Aside from differences in cropping the
projection data, the algorithm parameters are the same for each of the three DBT
data sets.  

Recall that the goal in designing the current ASD-POCS algorithm is to be able
to obtain images, within a few iterations (on the order of 10), corresponding
to any point in as-much-as-possible of allowed region of the data error-TV plane.
Starting with the microcalcification case, at the top of Fig. \ref{fig:traj},
the difference between the ASD-POCS and the standard EM algorithm is clear.
Reducing the value of $\beta$ seems to directly reduce image TV, and the 
adaptive component of the ASD-POCS allows the data error to be reduced with little
change in image TV.  The last iteration shown, number 20, at the bottom
of each of the three $\beta$ curves, is the minimum data-error image in the sequence.
Interestingly, this minimum data-error value seems to have little dependence on $\beta$
even though the image TV is dramatically reduced by lowering $\beta$. This is not
surprising due to the fact that the DBT system is very much undersampled in the
angular direction; many images with very different TV-values may correspond to the same
data-error. The track of the EM algorithm shows the traditional trade-off for most
iterative algorithms. As iteration number is increased data-error is reduced at the
expense of image regularity. For this particular EM run, no TV regularization was used.
But incorporating such regularization in EM, for example by the method discussed in
Refs. \cite{Lange:90,PerssonLimited:01}, results in an iteration track of similar shape. It is still
difficult to obtain images for the low-data-error, low-TV corner with a non-adaptive
iterative algorithm.  We point out that the ASD-POCS algorithm likely cannot explore
the complete allowed region of the data error-TV plane, especially within a few iterations.
And there is room for further algorithm development in pushing toward low-data-error and
low-image-TV.

Turning to the DBT case with the uniform mass, shown in the middle graph in Fig. \ref{fig:traj},
the algorithm trajectories are similar to the previous case aside from one aspect.
There is a significant drop in data error obtained by reducing $\beta$ from 1.0 to 0.1 .
This trend is counter-intuitive, because greater image regularity is generally obtained
at the expense of data fidelity.  In this case, imposing greater image regularity allows
for greater progress in reducing data-error.  This type of behavior, we have observed before
in image-reconstruction from simulated data; it generally occurs when the primary component
of the data error is noise in the detector bin measurements.  The data for this case is
noisier than that of the previous, microcalcification case. This is seen in the reconstructed
images, and the raw projection data show higher X-ray attenuation.  Yet, the minimum data-error
reached, at $\beta=0.1$, is comparable to minimum values reached for the microcalcification case.

Examining the curves for the spiculated-mass case, the shape of the curves is similar to that
of the microcalcification case. The difference between this case and the previous two is the
value of the minimum data-error achieved. It is roughly a factor of two higher than the previous
cases. Again, as this is a dense breast, the data noise is relatively high. But as $\beta$ is decreased
the data error remains high.  We speculate that the reason for this is that there may be additional
error due to incorrect geometry, such as patient motion during the scan.

Studying the algorithm trajectories in the data-error, image-regularity plane helps to
understand the image-reconstruction algorithm. Such curves may also prove useful in determining
data quality.  Clearly, for ideal data, a data-error of zero can be reached.  Data-error values, however,
will in general be finite, but it may be also important to know the source of the data inconsistency.
If these curves can be used to reveal data-error due to patient motion, they have additional, practical value.
For example, imaging microcalcifications is highly dependent on the absence of motion.
If a particular scan reveals no microcalcifications and the algorithm trajectories suggest patient motion
is likely present, it may be advisable to do a re-scan.

\section{discussion}
\label{sec:discussion}

We have introduced a practical, iterative image-reconstruction algorithm, within the ASD-POCS framework,
that can achieve useful images within a few iterations. This algorithm allows for fine control
over the regularity of the reconstructed images, which is essential for under-determined imaging
problems such as DBT. For the studies presented here, the image regularity metric is taken to be
the total $p$-variation, which reduces to the total variation and the image roughness for $p=1.0$
and $p=2.0$, respectively. The other main algorithm parameter, $\beta$, controls the level of the
regularity objective-function. As with all other iterative algorithms, the iteration number is
implicitly another parameter. The main advantage of the present algorithm is that each of these
few parameters have a real effect on the image quality, and these effects are relatively independent
of each other.

For DBT imaging, microcalcification imaging is the task that appears to be most greatly
impacted by the present algorithm.  Images reconstructed with low values of $p$ show markedly
greater contrast of the microcalcifications than those reconstructed by existing algorithms.
The practical significance of this increased contrast is that it may be possible to reduce
the X-ray intensity thereby lowering patient dose for the DBT scan.  The effects for mass
imaging are more subtle, but the finer controls allowed by the present algorithm
may allow better optimization of the DBT system for mass imaging by either human or computer
observers.

Extensions of this work can follow many different paths.  Within the ASD-POCS framework, various
methods of performing the adaptive control may lead to more efficient image-reconstruction
algorithms. Also different objective functions, which can simply be dropped into the present
framework, may be advantageous for different imaging tasks.  One practical question that
we intend to investigate is to use the ASD-POCS framework together with algorithm trajectories
to provide an assessment of projection data quality, particularly, to find a way to automatically
detect patient motion.

We point out that the algorithm presented here, though applied to DBT imaging, can easily
be adapted to other X-ray based tomographic systems. In fact, other tomographic imaging
modalities with a linear data model may also be amenable to image-reconstruction
within the ASD-POCS framework.

\begin{acknowledgments}

EYS and XP were supported in part by NIH R01 grants CA120540 and
EB000225, and by an Illinois Department of Public Health Ticket
for the Cure Grant.
ISR and RMN were supported in part by NIH grant  R33 CA109963 and R21 EB8801.
The original MGH-GE instrument was funded by the US ARMY via Clinical
translational research (CTR)DAMD17-98-8309, GE provided the current
clinical prototype system designated DBT Senographe DS.
NIH-NCI funded the use of this instument to aquire 3000 screening studies as 5R33CA107863-01.
Computations for this work were performed on a cluster,
partially funded by grants NIH S10 RR021039
and P30 CA14599.
The contents of this article are
solely the responsibility of the authors and do not necessarily
represent the official views of the National Institutes of Health.
\end{acknowledgments}

\bibliographystyle{apsrev}
%\bibliography{strings,refs,manuals}
\bibliography{tomo,sparse,tv-sidky}

\end{document}